\documentclass{emulateapj}
\usepackage{graphicx}
\usepackage{natbib}

\def\kms{km~s$^{-1}$}
\def\ergss{erg~s$^{-1}$}
\def\Geff{$\Gamma_{\rm eff}$}

\shorttitle{THE OPTX PROJECT III}
\shortauthors{Trouille et al.}

\begin{document}

\title{THE OPTX PROJECT III: X-RAY VERSUS OPTICAL SPECTRAL TYPE FOR AGNS\altaffilmark{1}}

\author{L. Trouille\altaffilmark{2}, A. J. Barger\altaffilmark{2,3,4},
  L. L. Cowie\altaffilmark{4}, Y. Yang\altaffilmark{5}, and
 R. F. Mushotzky\altaffilmark{6}}

\altaffiltext{1}{Some of the data presented herein were obtained at the W. M. Keck Observatory, which is operated as a scientific partnership among the California Institute of Technology, the University of California, and the National Aeronautics and Space Administration. The observatory was made possible by the generous financial support of the W. M. Keck Foundation.}
\altaffiltext{2}{Department of Astronomy, University of Wisconsin-Madison, 475 N. Charter Street, Madison, WI 53706}
\altaffiltext{3}{Department of Physics and Astronomy, University of Hawaii, 2505 Correa Road, Honolulu,
 HI 96822}
\altaffiltext{4}{Institute for Astronomy, University of Hawaii, 2680 Woodlawn Drive, Honolulu, HI 96822}
\altaffiltext{5}{Department of Astronomy, University of Illinois, 1002 W. Green St., Urbana, IL 61801}
\altaffiltext{6}{NASA Goddard Space Flight Center, Code 662, Greenbelt, MD 20771}

\begin{abstract}

We compare the optical spectral types with the X-ray spectral properties
for a uniformly selected (sources with fluxes greater than the
$3~\sigma$ level and above a
flux limit of $f_{2-8~\rm keV}>3.5\times 10^{-15}$~erg~cm$^{-2}$~s$^{-1}$),
highly spectroscopically complete ($>80$\% for
$f_{2-8~\rm keV}>10^{-14}$~erg~cm$^{-2}$~s$^{-1}$ and 
$>60$\% below) $2-8$~keV X-ray sample observed in three \emph{Chandra\/} 
fields (CLANS, CLASXS, and the CDF-N) that cover $\sim 1.2$~deg$^2$.
For our sample of 645 spectroscopically observed sources, 
we confirm that there is significant overlap
of the X-ray spectral properties, as determined by the effective
photon indices, \Geff, obtained from the ratios of the $0.5-2$~keV
to $2-8$~keV counts, for the different optical
spectral types.  For example, of the broad-line AGNs (non-broad-line 
AGNs), 20\%$\pm 3$\% (33\%$\pm 4$\%) have \Geff$<1.2$ (\Geff$\ge1.2$). 
Thus, one cannot use the X-ray spectral classifications and the 
optical spectral classifications equivalently. Since it is not
understood how X-ray and optical classifications relate to the
obscuration of the central engine, we strongly advise against a mixed
classification scheme, as it can only complicate the interpretation of
X-ray AGN samples. We confirm the dependence 
of optical spectral type on X-ray luminosity, and for $z<1$, we find a 
similar luminosity dependence of \Geff.  However, this dependence
breaks down at higher redshifts due to the highly redshift-dependent 
nature of \Geff.  We therefore also caution that any classification
scheme which depends on \Geff~is likely to suffer from serious
redshift bias.

\end{abstract}

\keywords{cosmology: observations  --- galaxies: active}

\section{Introduction}
\label{intro} 

Starting with the first \emph{Chandra\/} observations which resolved
the X-ray background \citep{mushotzky00}, ultradeep, small-area
(2~Ms CDF-N, \citealt{brandt01}, \citealt{alexander03}; 2~Ms CDF-S,
\citealt{giacconi02}, \citealt{luo08}) and intermediate-depth, wider-area
(SEXSI, \citealt{harrison03}; CLASXS, \citealt{yang04}; 
AEGIS-X, \citealt{nandra05}, \citealt{laird09}; 
extended-CDF-S or eCDF-S, \citealt{lehmer05}, \citealt{virani06}; 
ChaMP, \citealt{kim07a}; CLANS, \citealt{trouille08};
COSMOS, \citealt{elvis09}) {\em Chandra\/} X-ray surveys have 
uncovered a substantial population of active galactic nuclei (AGNs) 
that were not previously identified in optical or soft X-ray surveys, 
revolutionizing our understanding of accretion onto supermassive 
black holes.  Spectroscopic follow-up is essential for tracing the 
evolution of X-ray-selected AGNs over cosmic time, and over the years
a large number of redshifts have been obtained for all of the above
fields. However, the CDF-N
(\citealt{hornschemeier01}; \citealt{barger02}; \citealt{barger03};
\citealt{trouille08}), 1~Ms CDF-S (\citealt{szokoly04}),
CLASXS (\citealt{steffen04}; \citealt{trouille08}), and
CLANS (\citealt{trouille08}) fields are the most uniformly
spectroscopically complete of all of the {\em Chandra\/} surveys to date,
and the high-quality spectral data in these fields can be used to 
classify the sources optically.
(See Table~1 in \citealt{trouille08} for a summary of the 
spectroscopic completeness of the eCDF-S, AEGIS-X, SEXSI, 
and ChaMP fields and Figure~\ref{hfrac} in this paper for 
the combined completeness of our CDF-N, CLASXS, and CLANS fields.)

This is the third paper in our OPTX series, which focuses on the 
analysis of the X-ray sources in the CDF-N, CLASXS, and CLANS
fields. In the first paper (\citealt{trouille08}) we presented 
a new X-ray catalog for the CLANS field, as well as new (CLANS) 
and updated (CLASXS, CDF-N) redshift catalogs for the three fields. 
In the second paper (\citealt{yencho09}) we constructed rest-frame 
hard X-ray luminosity functions using our three fields, the CDF-S, 
an ASCA survey (\citealt{akiyama00}), and the local SWIFT 9-month
Burst Alert Telescope (BAT) survey 
(\citealt{tueller08}; \citealt{winter08, winter09}).  
Since it is very important to have a physically motivated 
classification scheme for understanding AGN evolution,
in this paper we use our high-quality Deep Imaging Multi-Object 
Spectrograph \citep[DEIMOS;][]{faber03} data from Keck~II to 
explore the differences between the optical, X-ray, and mixed 
classification schemes that have been proposed for analyzing 
{\em Chandra\/} X-ray samples.  

As two different groups began carrying out extensive spectroscopy of 
the ultradeep CDFs and releasing the measurements to the public, it 
became clear that a consistent optical classification 
system for these faint X-ray sources was needed so that the data 
could be analyzed together. 
\citet{szokoly04} and \citet{barger05} pointed out that it would
be problematic to apply classical optical AGN definitions to the 
faint X-ray sources due to the range of rest-frame wavelengths 
covered and the varying degree of mixing of the AGN spectrum with 
the host galaxy spectrum at different redshifts.  Moreover, by this 
time it was already well known that many luminous X-ray sources 
showed no signatures of AGN activity in their optical spectra 
(e.g., \citealt{barger01}; \citealt{hornschemeier01}).  
Thus, \citet{szokoly04} proposed a new optical 
classification scheme, which \citet{barger05} roughly matched 
by defining the following spectral types: 
(1) absorbers (ABS; no strong emission lines); 
(2) star formers (SF; strong Balmer lines and no broad or 
high-ionization lines); 
(3) high-excitation sources (HEX; [NeV], CIV, narrow MgII lines, or 
strong [OIII]); 
and (4) broad-line AGNs (BLAGNs; optical lines having 
FWHM line widths $>2000$~km~s$^{-1}$).  Although
the HEX spectral type largely overlaps the classical Seyfert~2 
spectral type, describing the sources as HEX sources helps to 
avoid confusion with the classical definitions.  

X-ray data alone have also long been used to estimate the 
amount of obscuration between the observer and the nuclear 
source through the $0.5-8~\rm{keV}$ spectral slope. 
Since $2-8~\rm{keV}$ X-rays will penetrate obscuring
material (except in Compton-thick AGNs, where the neutral 
hydrogen column density, $N_{\rm H}$, in the line of sight is 
higher than the inverse Thomson cross section, 
$N_{\rm H}\simeq 1.5\times 10^{24}$~cm$^{-2}$) and
$0.5-2~\rm{keV}$ X-rays will not, in low signal-to-noise data a
shallower slope may indicate an
obscured source.  X-ray spectra can be approximated with a power-law of
the form $P(E) = AE^{-\Gamma_{\rm eff}}$, where $E$ is the photon
energy in $\rm{keV}$ and $A$ is the normalization factor. 
(Here we indicate the power law slope as an \emph{effective} $\Gamma$ 
to distinguish it from the intrinsic slope, $\Gamma$. 
\Geff~is not the true $\Gamma$ unless there is no intrinsic 
absorption.) 
\citet{barger05} showed that \citet{szokoly04}'s choice of
2000~km~s$^{-1}$ as the dividing line between BLAGNs and non-BLAGNs 
in the optical classification scheme made sense in terms of
the X-ray spectral properties,
as above 2000~km~s$^{-1}$ almost all of the
sources are X-ray soft ($\Gamma_{eff}=1.8$), whereas below this 
line width there is a wide span of X-ray colors.

In addition to their optical classification scheme,
\citet{szokoly04} introduced a classification 
scheme that follows the unified model for AGNs (\citealt{antonucci93})
and is based solely on the X-ray properties of the sources.  
In this scheme objects with unabsorbed X-ray luminosities stronger 
than expected from stellar processes in normal galaxies 
(i.e., $L_{0.5-10~{\rm keV}}\ge 10^{42}$~erg~s$^{-1}$)
are classified as AGNs, and unabsorbed sources are separated
from absorbed sources through the use of a {\em Chandra\/}-specific
X-ray hardness ratio, HR=($C_{hard}-C_{soft}$)/($C_{hard} +C_{soft}$), 
where $C_{hard}$ and $C_{soft}$ are the net ACIS-I count rates in the 
hard ($2-10$~keV) and soft ($0.5-2$~keV) bands, respectively. 
They chose HR $\le -0.2$ to indicate an unabsorbed source
(which they call an X-ray type~1 source) and HR $>-0.2$ to indicate
an absorbed sources (which they call an X-ray type~2 source).  
(Note that this HR is approximately equivalent to, at $z=0$, our 
$\Gamma_{\rm eff}=1.2$.)
However, they noted that while an increasing absorption makes a 
source harder, a higher redshift makes a source softer (the $2-10~\rm
keV$ filter samples higher energies which are less affected by
obscuring material), which means this approach might
incorrectly identify a high-redshift absorbed (X-ray type~2) source as
an unabsorbed (X-ray type~1) source.

\citet{hasinger05} took the classification of
X-ray sources one stop further, arguing for a third scheme 
that is a mix of the above two schemes. 
This decision was based on their assumption that `true' 
BLAGNs may be optically misclassified
as non-BLAGNs due to dilution of the AGN light by the host
galaxy light, particularly at lower luminosities 
(\citealt{moran02,severgnini03,garcet07,cardamone07}).
Thus, according to \citet{hasinger05}, the most 
`complete' sample of unobscured AGNs is one which includes any object 
optically classified as a BLAGN, as well as any object satisfying 
$L_{0.5-10~\rm keV}\ge 10^{42}$~erg~s$^{-1}$ and HR $\le -0.2$.

Unfortunately, the creation of a classification scheme that mixes optical 
and X-ray spectral diagnostics requires a thorough understanding of the
correspondence betweeen X-ray and optical spectral type. 
However, it is well known observationally (yet unexplained physically)
that $\sim 10-30$\% of AGNs have (1) X-ray spectra that show no absorption
and (2) optical spectra that suggest obscuration 
(e.g., \citealt{pappa01}; \citealt{panessa02}; \citealt{barcons03};
\citealt{georgantopoulos03}; \citealt{caccianiga04}; \citealt{corral05}; 
\citealt{wolter05}; \citealt{mateos05}; \citealt{tozzi06}). 
The opposite effect, i.e., (1) X-ray spectra that show absorption
and (2) optical spectra that suggest no obscuration, has also been 
observed (e.g., \citealt{silverman05} found that $\sim 15$\% of X-ray 
hard AGNs in ChaMP are BLAGNs; other examples can be found in
\citealt{comastri01}; \citealt{wilkes02}; 
\citealt{fiore03}; \citealt{brusa03}; \citealt{akiyama03};
\citealt{silverman05}; \citealt{gallagher06}; \citealt{hall06}; 
\citealt{tajer07}).

Moreover, the assumption that galaxy dilution is causing BLAGNs to be
optically misclassified as non-BLAGNs has been 
called into question by two studies.  
\citet{barger05} used the \emph{Hubble Space Telescope} 
Great Observatories Origins Deep Survey North (GOODS-N; 
\citealt{giavalisco04}) 
data to measure the nuclear UV magnitudes of AGNs in the CDF-N.  It is 
well known that
the nuclear UV magnitudes of BLAGNs are strongly correlated with the
$0.5-2$~keV fluxes (e.g., \citealt{zamorani81}), and, indeed 
\citet{barger05} found that their BLAGNs also showed this correlation.
However, they also found that the non-BLAGNs in their sample did not; 
rather, the UV nuclei of these sources were much weaker relative to 
their X-ray light than would be expected if they were similar to the BLAGNs.
In the second study, \citet{cowie09} combined Galaxy Evolution Explorer
(GALEX; \citealt{martin05}) data with the CDF-N, CLASXS, and CLANS
X-ray samples to determine the ionizing flux from $z\sim 1$ AGNs.
They found that while the BLAGNs in their sample exhibited substantial UV
ionizing flux, the non-BLAGNs were UV faint.

Here we use our extensive observations of the CDF-N, CLASXS, and 
CLANS fields to analyze the optical and X-ray spectral properties 
of a significant (fluxes $>3\sigma$ level), intermediate-depth
($f_{2-8~\rm keV}>3.5\times 10^{-15}$~erg~cm$^{-2}$~s$^{-1}$) sample 
of X-ray sources.  Due to the need for a large and 
very complete optical spectroscopic sample, this is the first time 
that such a comparative study has been done. 
Of particular interest is the relationship between the 
X-ray spectral properties and the optical spectral properties of 
X-ray-selected sources and whether such characteristics map each 
other well enough that they can be merged into a single 
classification scheme.

The structure of the paper is as follows. In Section~\ref{xray} we 
describe our X-ray samples. In Section~\ref{optspec} we discuss our 
spectroscopic completeness and optical classifications and show the 
redshift distributions by optical spectral type. 
In Section~\ref{gamma} we compare our optical 
classifications with the X-ray classifications, and we illustrate 
the challenges of a mixed classification scheme.
In Section~\ref{discussion} we examine in the context of our
data the observational obstacles that have been invoked to explain 
the observed mismatches between X-ray and optical spectral types, and
in Section \ref{conclusions} we present our conclusions. 

All magnitudes are in the Vega magnitude system. We assume
$\Omega_M=0.3, \Omega_{\Lambda}=0.7$, and 
H$_0=70$~km~s$^{-1}$~Mpc$^{-1}$ throughout.

\section{X-ray Data}
\label{xray}

We construct our $2-8~\rm keV$ sample from three different fields in the sky, 
so known field-to-field variations due to large scale structure are
minimized. All three of our fields sample regions of low Galactic HI 
column density. Both the CLANS and CLASXS fields reside in the Lockman 
Hole high-latitude region of extremely low Galactic HI column density
\citep[$5.7\times10^{19}$ cm$^{-2}$;][]{lockman86}. The Galactic HI
column density along the line of sight to the CDF-N is
$1.6\times10^{20}$ cm$^{-2}$ \citep{stark92}.

\begin{table}
\begin{small}
\centering
\caption{CLANS, CLASXS, and CDF-N Survey Characteristics}
\label{surveys}
\begin{tabular}{l c c c}
\tableline\tableline
   & CLANS & CLASXS & CDF-N \\
\tableline
Total Exposure   &  70~ks &  40~ks (70 ks\tablenotemark{b}) & 2~Ms  \\
Area (deg$^2$)         &  0.6       &  0.45        & 0.124        \\
$2-8~\rm keV$ flux limit\tablenotemark{a} &   35     & 60 (35\tablenotemark{b})    & 1.4       \\
\tableline
\end{tabular}
\end{small}
\footnotesize
\tablenotetext{a}{$10^{-16}$~erg~cm$^{-2}$~s$^{-1}$; $3~\sigma$ flux limit
at the pointing center, see \citet{trouille08}.}
\tablenotetext{b}{The central CLASXS pointing.}
\end{table}

In Table~\ref{surveys} we list the \emph{Chandra} exposure times, the
areas covered, 
and the $2-8$~keV $3~\sigma$ flux limits at the pointing centers.
Because the CLANS and CLASXS fields are shallower than the CDF-N field, 
we have limited our study to sources with fluxes greater than the
$3~\sigma$ flux limit at the pointing center of the $70$~ks CLANS 
pointings 
(i.e., $f_{2-8~\rm keV}>3.5\times 10^{-15}$~erg~cm$^{-2}$~s$^{-1}$).
We note that although only the central CLASXS pointing is
$\sim70$~ks (the other eight are $40$~ks), the CLASXS survey was
designed to achieve uniform field coverage, and so there is
substantial overlap between pointings.  Also, since our goal is simply
to create a uniform sample of sources above a given flux limit for the
purposes of studying their X-ray and optical spectral characteristics,
the choice of flux limit will not greatly affect our present results 
(as opposed to, for example, the impact it would have on determining 
the number densities). 

To construct a significant $2-8~\rm keV$ sample,
we additionally only use sources with fluxes greater than the 
$3~\sigma$ level.  We determined this by using the $1~\sigma$ error 
bars on the $2-8$~keV fluxes given in \citet{alexander03},
\citet{yang04}, and \citet{trouille08} for the CDF-N, CLASXS, and
CLANS fields, respectively.  Hereafter, this paper's ``$2-8$~keV sample'' 
consists of 745 X-ray sources selected in the $2-8$~keV band.

\section{Classification by Optical Spectral Type}
\label{optspec}

\begin{table*}
\begin{tiny}
\centering
\caption{CLANS Catalog, Updated Sources}
\label{clansupdate}
\begin{tabular}{lccccccccc}
\tableline\tableline
Num. & $\alpha_{2000}$ & $\delta_{2000}$ & n$_{0.5-2~\rm keV}$ &
n$_{2-8~\rm keV}$ & $f_{0.5-2~\rm keV}$$^a$ &
$f_{2-8~\rm keV}$$^a$ & $f_{0.5-8~\rm keV}$$^a$ & $z_{spec}$ & class \\
(1) & (2) & (3) & (4) & (5) & (6) & (7) & (8) & (9) & (10)\\
\tableline
 80 & 160.8983 & 58.9781 &   0.00$_{-  0.00}^{+  1.83}$ &   8.25$_{-  3.03}^{+  3.69}$ &   0.00$_{-  0.00}^{+  0.03}$ &   2.10$_{- 32.50}^{+ 39.51}$ &   2.03$_{-  1.01}^{+  1.01}$ &   1.30 &   1\\
150 & 161.0324 & 58.6700 &   0.00$_{-  0.00}^{+  1.83}$ &  12.57$_{-  3.14}^{+  5.12}$ &   0.00$_{-  0.00}^{+  0.03}$ &   2.70$_{- 29.88}^{+ 48.70}$ &   2.22$_{-  1.11}^{+  1.11}$ &   0.00 & -99\\
163 & 161.0519 & 58.5388 &  70.75$_{-  8.16}^{+  9.71}$ &  37.75$_{-  5.89}^{+  7.47}$ &   7.11$_{-  0.82}^{+  0.98}$ &  15.90$_{-  2.48}^{+  3.14}$ &  23.00$_{-  2.18}^{+  2.52}$ &   0.97 &   3\\
380 & 161.4857 & 58.9907 &   4.00$_{-  1.94}^{+  3.15}$ &   5.75$_{-  2.15}^{+  3.83}$ &   0.34$_{-  0.17}^{+  0.27}$ &   2.57$_{-  0.96}^{+  1.71}$ &   1.82$_{-  0.77}^{+  0.97}$ &   1.10 &   1\\
381 & 161.4875 & 59.0001 &  15.00$_{-  3.84}^{+  4.95}$ &   5.50$_{-  1.90}^{+  4.08}$ &   2.61$_{-  0.67}^{+  0.86}$ &   3.35$_{-  1.16}^{+  2.49}$ &   6.43$_{-  1.30}^{+  1.78}$ &   3.38 &   3\\
390 & 161.5037 & 59.1229 &  68.75$_{-  8.04}^{+  9.59}$ &  35.75$_{-  5.73}^{+  7.30}$ &   6.81$_{-  0.80}^{+  0.95}$ &  14.51$_{-  2.33}^{+  2.96}$ &  21.02$_{-  2.00}^{+  2.42}$ &   1.22 &   3\\
404 & 161.5408 & 58.9262 &   0.00$_{-  0.00}^{+  1.83}$ &  43.25$_{-  6.79}^{+  7.36}$ &   0.00$_{-  0.00}^{+  0.02}$ &   9.90$_{- 94.45}^{+102.34}$ &   8.02$_{-  4.01}^{+  4.01}$ &   0.29 &   2\\
426 & 161.5699 & 58.6507 &  40.00$_{-  6.30}^{+  7.37}$ &  19.50$_{-  4.83}^{+  4.93}$ &   4.22$_{-  0.67}^{+  0.78}$ &   8.32$_{-  2.06}^{+  2.10}$ &  13.01$_{-  1.72}^{+  1.84}$ &   0.68 &   2\\
431 & 161.5748 & 58.6362 &  84.50$_{-  8.71}^{+ 10.75}$ &  51.50$_{-  6.69}^{+  8.75}$ &   7.61$_{-  0.78}^{+  0.97}$ &  20.04$_{-  2.61}^{+  3.41}$ &  27.92$_{-  2.46}^{+  2.57}$ &   0.39 &   4\\
440 & 161.5923 & 58.6561 &  25.50$_{-  4.57}^{+  6.66}$ &  13.25$_{-  3.82}^{+  4.44}$ &   2.78$_{-  0.50}^{+  0.72}$ &   5.56$_{-  1.60}^{+  1.86}$ &   8.53$_{-  1.41}^{+  1.53}$ &   0.95 &   3\\
441 & 161.5930 & 58.7007 &  11.00$_{-  3.28}^{+  4.41}$ &  23.75$_{-  4.62}^{+  6.21}$ &   0.84$_{-  0.25}^{+  0.34}$ &  10.81$_{-  2.10}^{+  2.83}$ &   8.77$_{-  1.52}^{+  2.18}$ &   0.98 &   1\\
464 & 161.6507 & 58.6700 &  14.00$_{-  3.71}^{+  4.83}$ &   5.00$_{-  2.18}^{+  3.37}$ &   1.30$_{-  0.34}^{+  0.45}$ &   1.67$_{-  0.73}^{+  1.13}$ &   2.84$_{-  0.63}^{+  0.89}$ &   1.07 &   3\\
468 & 161.6623 & 59.2879 &  17.75$_{-  3.96}^{+  5.57}$ &  12.50$_{-  3.07}^{+  5.19}$ &   1.57$_{-  0.35}^{+  0.49}$ &   4.81$_{-  1.18}^{+  2.00}$ &   6.73$_{-  1.20}^{+  1.44}$ &   0.16 &   1\\
472 & 161.6673 & 58.6292 &  20.50$_{-  4.06}^{+  6.15}$ &  16.50$_{-  4.47}^{+  4.58}$ &   1.79$_{-  0.35}^{+  0.54}$ &   6.76$_{-  1.83}^{+  1.88}$ &   8.46$_{-  1.41}^{+  1.66}$ &   2.07 &   3\\
509 & 161.7208 & 59.3966 &   1.25$_{-  1.12}^{+  2.04}$ &  15.50$_{-  3.47}^{+  5.58}$ &   0.09$_{-  0.08}^{+  0.15}$ &  11.02$_{-  2.47}^{+  3.97}$ &   8.98$_{-  2.26}^{+  3.31}$ &   1.22 &   0\\
523 & 161.7517 & 59.3461 &   3.50$_{-  1.44}^{+  3.65}$ &  19.75$_{-  4.19}^{+  5.79}$ &   0.26$_{-  0.11}^{+  0.27}$ &  11.42$_{-  2.42}^{+  3.35}$ &   5.77$_{-  1.73}^{+  2.03}$ &   5.29 &   1\\
558 & 161.7836 & 58.6432 &  22.25$_{-  4.91}^{+  5.51}$ &  32.75$_{-  5.47}^{+  7.05}$ &   1.88$_{-  0.41}^{+  0.46}$ &  15.53$_{-  2.60}^{+  3.34}$ &  14.70$_{-  1.98}^{+  2.62}$ &   2.44 &   3\\
561 & 161.7889 & 58.6454 &  36.25$_{-  6.23}^{+  6.80}$ &  30.00$_{-  5.45}^{+  6.53}$ &   3.18$_{-  0.55}^{+  0.60}$ &  12.67$_{-  2.30}^{+  2.76}$ &  13.93$_{-  1.72}^{+  2.22}$ &   3.17 &   3\\
634 & 161.9402 & 59.4728 &  20.50$_{-  4.06}^{+  6.15}$ &  34.75$_{-  5.64}^{+  7.22}$ &   1.90$_{-  0.38}^{+  0.57}$ &  18.47$_{-  3.00}^{+  3.84}$ &  19.36$_{-  2.71}^{+  2.91}$ &   0.83 &   2\\
636 & 161.9435 & 59.3978 &   2.00$_{-  1.32}^{+  2.63}$ &   9.50$_{-  3.46}^{+  3.60}$ &   0.16$_{-  0.10}^{+  0.21}$ &   5.56$_{-  2.03}^{+  2.11}$ &   4.89$_{-  1.29}^{+  2.29}$ &  -2.00 &  -2\\
645 & 161.9531 & 59.3914 &  97.75$_{-  9.64}^{+ 11.18}$ &  48.75$_{-  6.73}^{+  8.29}$ &   9.54$_{-  0.94}^{+  1.09}$ &  19.26$_{-  2.66}^{+  3.28}$ &  27.97$_{-  2.32}^{+  2.63}$ &   1.56 &   4\\
712 & 162.0803 & 58.7750 &   6.75$_{-  2.35}^{+  4.01}$ &  20.00$_{-  4.44}^{+  5.54}$ &   0.52$_{-  0.18}^{+  0.31}$ &   9.66$_{-  2.15}^{+  2.68}$ &  10.95$_{-  1.88}^{+  2.45}$ &   0.58 &   2\\
733 & 162.1368 & 59.2878 &  74.75$_{-  8.40}^{+  9.95}$ &  44.75$_{-  6.44}^{+  8.01}$ &   7.53$_{-  0.85}^{+  1.00}$ &  19.25$_{-  2.77}^{+  3.44}$ &  26.93$_{-  2.42}^{+  2.78}$ &   1.11 &   1\\
756 & 162.2156 & 58.7401 &   5.00$_{-  2.18}^{+  3.37}$ &   6.00$_{-  2.40}^{+  3.58}$ &   1.63$_{-  0.41}^{+  0.47}$ &  13.61$_{-  2.55}^{+  3.39}$ &  14.46$_{-  2.30}^{+  2.49}$ &   0.53 &   2\\
761 & 162.2203 & 58.9016 &  49.25$_{-  7.23}^{+  7.79}$ &  28.50$_{-  4.86}^{+  6.94}$ &   4.38$_{-  0.64}^{+  0.69}$ &  10.33$_{-  1.76}^{+  2.52}$ &  14.39$_{-  1.70}^{+  1.81}$ &   1.13 &   4\\
762 & 162.2264 & 58.8028 &   9.00$_{-  2.96}^{+  4.10}$ &  16.00$_{-  3.97}^{+  5.08}$ &   0.74$_{-  0.24}^{+  0.34}$ &   7.03$_{-  1.74}^{+  2.23}$ &   4.89$_{-  1.27}^{+  1.45}$ &   0.00 & -99\\
763 & 162.2360 & 59.1422 &  12.25$_{-  3.68}^{+  4.30}$ &   8.00$_{-  2.78}^{+  3.94}$ &   1.53$_{-  0.46}^{+  0.54}$ &   4.45$_{-  1.55}^{+  2.19}$ &   5.73$_{-  1.25}^{+  1.74}$ &  -1.00 &  -1\\
764 & 162.2364 & 58.8562 &  12.75$_{-  3.32}^{+  4.94}$ &  15.50$_{-  3.47}^{+  5.58}$ &   0.34$_{-  0.18}^{+  0.23}$ &   2.42$_{-  0.90}^{+  1.61}$ &   2.16$_{-  0.75}^{+  1.06}$ &   0.00 & -99\\
765 & 162.2370 & 58.8560 &  16.00$_{-  3.97}^{+  5.08}$ &  20.00$_{-  4.44}^{+  5.54}$ &   1.27$_{-  0.32}^{+  0.40}$ &   7.83$_{-  1.74}^{+  2.17}$ &   7.66$_{-  1.27}^{+  1.79}$ &   0.00 & -99\\
766 & 162.2384 & 58.9408 &   4.50$_{-  1.68}^{+  3.87}$ &   2.50$_{-  1.16}^{+  3.41}$ &   0.47$_{-  0.19}^{+  0.28}$ &   5.04$_{-  1.50}^{+  2.02}$ &   3.28$_{-  0.92}^{+  1.42}$ &   0.00 & -99\\
767 & 162.2402 & 58.8784 &   8.75$_{-  2.71}^{+  4.35}$ &   5.75$_{-  2.15}^{+  3.83}$ &   0.72$_{-  0.22}^{+  0.36}$ &   2.02$_{-  0.75}^{+  1.34}$ &   2.28$_{-  0.65}^{+  0.87}$ &   0.00 & -99\\
768 & 162.2450 & 58.9788 &  94.25$_{-  9.93}^{+ 10.48}$ &  32.50$_{-  5.22}^{+  7.30}$ &   9.89$_{-  1.04}^{+  1.10}$ &  12.05$_{-  1.94}^{+  2.71}$ &  22.29$_{-  1.98}^{+  2.16}$ &   1.36 &   3\\
769 & 162.2452 & 59.1244 &  50.25$_{-  7.30}^{+  7.87}$ &  26.00$_{-  5.07}^{+  6.16}$ &   4.66$_{-  0.68}^{+  0.73}$ &   9.32$_{-  1.82}^{+  2.21}$ &  14.04$_{-  1.62}^{+  1.81}$ &  -1.00 &  -1\\
770 & 162.2460 & 58.8365 &  24.75$_{-  4.72}^{+  6.31}$ &  23.25$_{-  5.02}^{+  5.61}$ &   2.03$_{-  0.39}^{+  0.52}$ &   8.66$_{-  1.87}^{+  2.09}$ &  11.06$_{-  1.46}^{+  1.92}$ &   0.00 & -99\\
771 & 162.2533 & 59.0769 &  15.25$_{-  4.09}^{+  4.70}$ &   0.00$_{-  0.00}^{+ 21.95}$ &   1.37$_{-  0.37}^{+  0.42}$ &   0.00$_{- 10.32}^{+  0.00}$ &   4.73$_{-  1.27}^{+  1.46}$ &  -1.00 &  -1\\
772 & 162.2644 & 59.0021 &  15.00$_{-  3.84}^{+  4.95}$ &   6.00$_{-  2.40}^{+  3.58}$ &   0.63$_{-  0.21}^{+  0.34}$ &   3.31$_{-  1.22}^{+  1.48}$ &   2.90$_{-  0.78}^{+  1.18}$ &  -1.00 &  -1\\
773 & 162.2718 & 58.8679 &  22.00$_{-  4.66}^{+  5.76}$ &   7.50$_{-  2.28}^{+  4.44}$ &   2.62$_{-  0.56}^{+  0.69}$ &   2.79$_{-  0.85}^{+  1.65}$ &   4.91$_{-  0.85}^{+  1.22}$ &   3.44 &   4\\
774 & 162.2795 & 58.9679 &   0.50$_{-  0.37}^{+  2.79}$ &   4.75$_{-  1.93}^{+  3.62}$ &   0.04$_{-  0.03}^{+  0.22}$ &   3.22$_{-  1.31}^{+  2.46}$ &   1.36$_{-  0.95}^{+  1.43}$ &   0.75 &   3\\
775 & 162.2949 & 58.8574 &   9.75$_{-  2.87}^{+  4.51}$ &   8.50$_{-  2.46}^{+  4.60}$ &   0.83$_{-  0.24}^{+  0.38}$ &   3.22$_{-  0.93}^{+  1.75}$ &   3.96$_{-  0.88}^{+  1.24}$ &   0.99 &   1\\
776 & 162.2998 & 59.0320 &  51.25$_{-  7.37}^{+  7.94}$ &  17.50$_{-  3.71}^{+  5.82}$ &   5.54$_{-  0.80}^{+  0.86}$ &   6.63$_{-  1.41}^{+  2.20}$ &  12.19$_{-  1.53}^{+  1.63}$ &  -1.00 &  -1\\
777 & 162.3024 & 58.8895 &  18.00$_{-  4.21}^{+  5.32}$ &   6.50$_{-  2.10}^{+  4.26}$ &   1.68$_{-  0.39}^{+  0.50}$ &   2.14$_{-  0.69}^{+  1.41}$ &   3.83$_{-  0.81}^{+  0.90}$ &   0.00 & -99\\
778 & 162.3048 & 58.9967 &  17.25$_{-  4.34}^{+  4.95}$ &  25.75$_{-  4.82}^{+  6.41}$ &   0.54$_{-  0.17}^{+  0.35}$ &   5.76$_{-  1.73}^{+  2.02}$ &   5.51$_{-  1.39}^{+  1.58}$ &  -1.00 &  -1\\
779 & 162.3198 & 58.8593 &   7.25$_{-  2.85}^{+  3.51}$ &  13.00$_{-  3.57}^{+  4.69}$ &   0.03$_{-  0.02}^{+  0.19}$ &  15.50$_{-  3.21}^{+  4.97}$ &  17.38$_{-  3.53}^{+  4.80}$ &   0.00 & -99\\
780 & 162.3240 & 58.8417 &  11.25$_{-  3.53}^{+  4.16}$ &  21.25$_{-  4.81}^{+  5.40}$ &   0.97$_{-  0.30}^{+  0.36}$ &  10.14$_{-  2.29}^{+  2.58}$ &   9.87$_{-  1.86}^{+  2.05}$ &   0.00 & -99\\
781 & 162.3285 & 58.8992 &  37.25$_{-  6.31}^{+  6.88}$ &  12.75$_{-  3.32}^{+  4.94}$ &   3.89$_{-  0.66}^{+  0.72}$ &   4.66$_{-  1.21}^{+  1.81}$ &   8.61$_{-  1.18}^{+  1.45}$ &   0.21 &   1\\
782 & 162.3314 & 58.9806 &  43.75$_{-  6.36}^{+  7.93}$ &  27.50$_{-  5.67}^{+  5.76}$ &   4.57$_{-  0.67}^{+  0.83}$ &  12.53$_{-  2.59}^{+  2.62}$ &  17.20$_{-  1.90}^{+  2.39}$ &   1.79 &   4\\
783 & 162.3320 & 58.7839 & 114.25$_{- 10.92}^{+ 11.46}$ &  48.00$_{-  6.91}^{+  7.97}$ &  10.97$_{-  1.05}^{+  1.10}$ &  16.98$_{-  2.44}^{+  2.82}$ &  27.42$_{-  2.18}^{+  2.36}$ &   2.36 &   3\\
784 & 162.3330 & 58.7929 &   6.50$_{-  2.10}^{+  4.26}$ &  12.25$_{-  3.68}^{+  4.30}$ &   1.45$_{-  0.43}^{+  0.58}$ &   0.20$_{-  0.16}^{+  0.68}$ &   1.40$_{-  0.40}^{+  0.47}$ &   0.00 & -99\\
785 & 162.3517 & 58.7906 &  11.00$_{-  3.28}^{+  4.41}$ &   0.75$_{-  0.62}^{+  2.54}$ &   1.51$_{-  0.35}^{+  0.49}$ &   5.39$_{-  1.55}^{+  1.81}$ &   6.83$_{-  1.36}^{+  1.38}$ &   0.00 & -99\\
786 & 162.3532 & 58.8503 &   8.50$_{-  2.46}^{+  4.60}$ &   4.00$_{-  1.94}^{+  3.15}$ &   0.42$_{-  0.16}^{+  0.36}$ &   0.95$_{-  0.44}^{+  1.30}$ &   0.90$_{-  0.33}^{+  0.77}$ &   0.00 & -99\\
787 & 162.3717 & 58.8940 & 838.75$_{- 28.71}^{+ 30.23}$ & 239.75$_{- 15.23}^{+ 16.76}$ &  90.78$_{-  3.11}^{+  3.27}$ &  86.92$_{-  5.52}^{+  6.08}$ & 176.46$_{-  5.38}^{+  5.64}$ &   0.25 &   4\\
788 & 162.3815 & 58.8004 &  16.75$_{-  3.84}^{+  5.45}$ &  13.25$_{-  3.82}^{+  4.44}$ &   1.48$_{-  0.44}^{+  0.52}$ &   2.99$_{-  1.27}^{+  1.59}$ &   4.44$_{-  0.92}^{+  1.42}$ &  -1.00 &  -1\\
789 & 162.4150 & 58.8908 &  12.25$_{-  3.68}^{+  4.30}$ &   6.25$_{-  2.65}^{+  3.33}$ &   2.03$_{-  0.41}^{+  0.56}$ &   6.40$_{-  1.43}^{+  2.30}$ &   6.67$_{-  1.28}^{+  1.41}$ &  -1.00 &  -1\\
\tableline
\end{tabular}
\end{tiny}
\footnotesize
\tablecomments{  \\
$^a$Units of $10^{-15}$~erg~cm$^{-2}$~s$^{-1}$.\\
$z_{spec} = 0$ and corresponding class $= -99$, source spectroscopically observed but neither the redshift nor the class could be identified. \\
$z_{spec} = -1$ and corresponding class $= -1$, source not yet spectroscopically observed. \\
$z_{spec} = -2$ and corresponding class $= -2$, source is a star. \\
class $=0$, absorbers; class $=1$, star formers; class $=3$, high-excitation sources; class $=4$, BLAGNs.}
\end{table*}

\begin{table}
\begin{small}
\centering
\caption{Optical Spectral Classification for our $2-8$~keV Sample by Field
  and Optical Spectral Type}
\label{hclass table}
\begin{tabular}{l c c c c c c}
\tableline\tableline
Category & CLANS & CLASXS & CDF-N & All \\
                   & (407)        & (251)         & (87)           & (745)\\
\tableline
Observed                & 331        & 235         & 85           & 651\\
Identified              & 243        & 162         & 64           & 469\\
Unidentified            & 88         & 73          & 21           & 182\\
Unidentified \\
\,\, with phot-zs       & 51         & 43          & 18           & 112 \\
BLAGN\tablenotemark{a}  & 106 (43\%) & 77 (47\%)   & 24 (38\%)    & 207 (44\%)\\
HEX\tablenotemark{a}    & 72 (30\%)  & 35 (22\%)   & 11 (17\%)    & 118 (25\%)\\
SF\tablenotemark{a}     & 51 (21\%)  & 34 (21\%)   & 19 (30\%)    & 104 (22\%)\\
ABS\tablenotemark{a}    & 13 (5\%)   & 12 (7\%)    & 8  (13\%)    & 33 (7\%)\\
Star                    & 1          & 4           & 1            & 6\\
\tableline
\end{tabular}
\end{small}
\footnotesize
\tablenotetext{a}{The percentages refer to the percent of identified 
sources that exhibit the specified spectral type.}
\end{table}

\begin{figure}[!ht]
\epsscale{2}
\plottwo{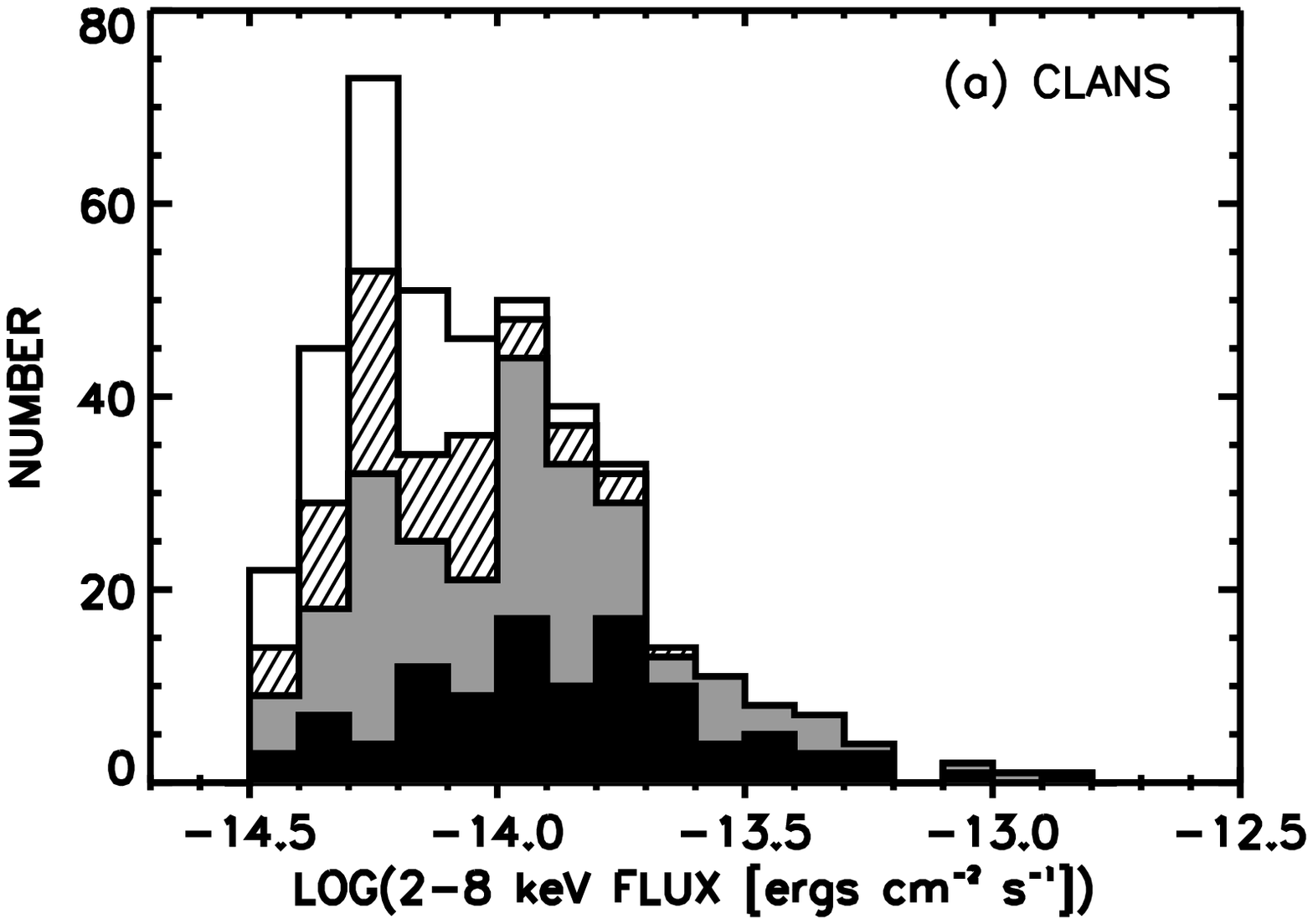}{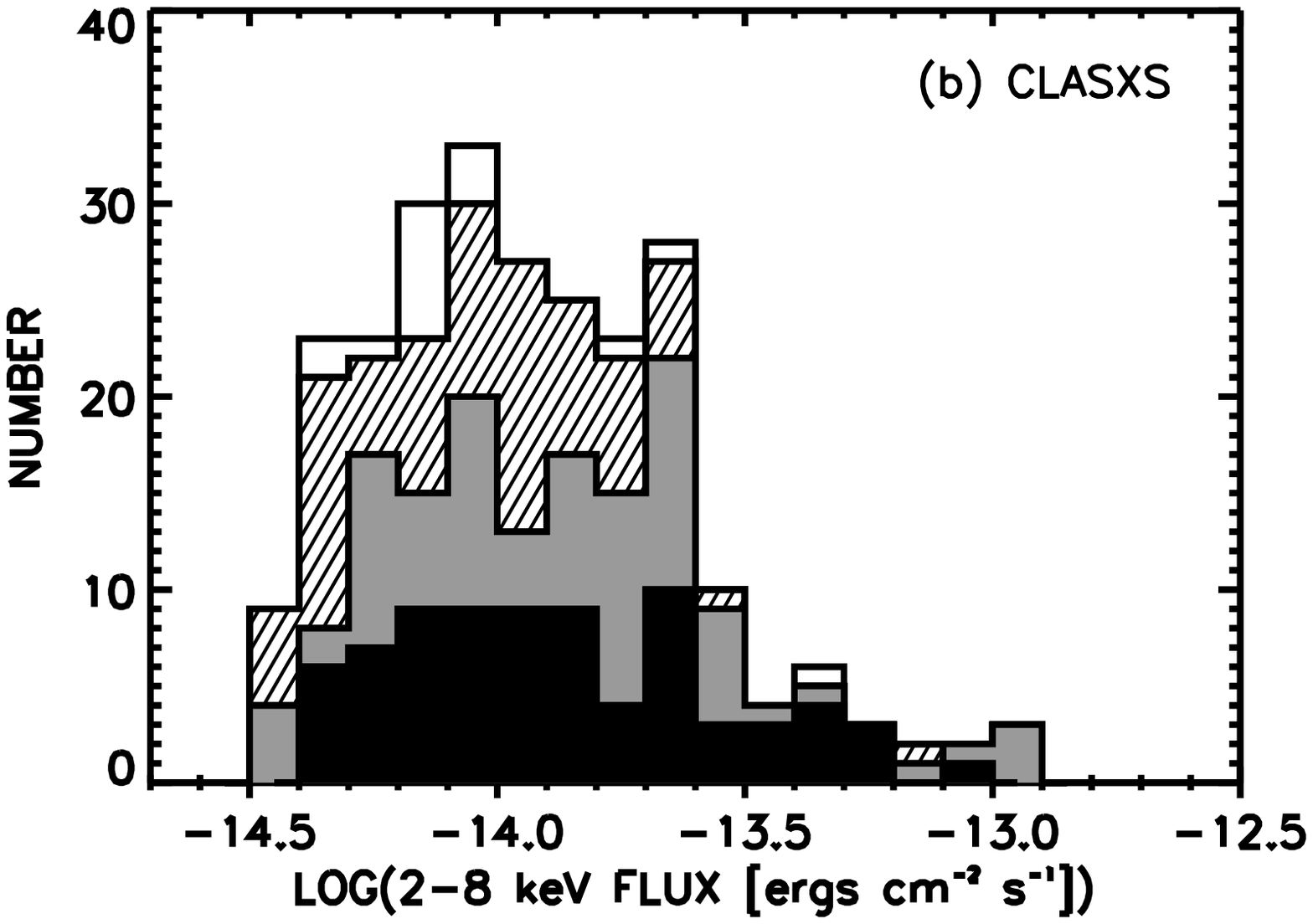}
\plottwo{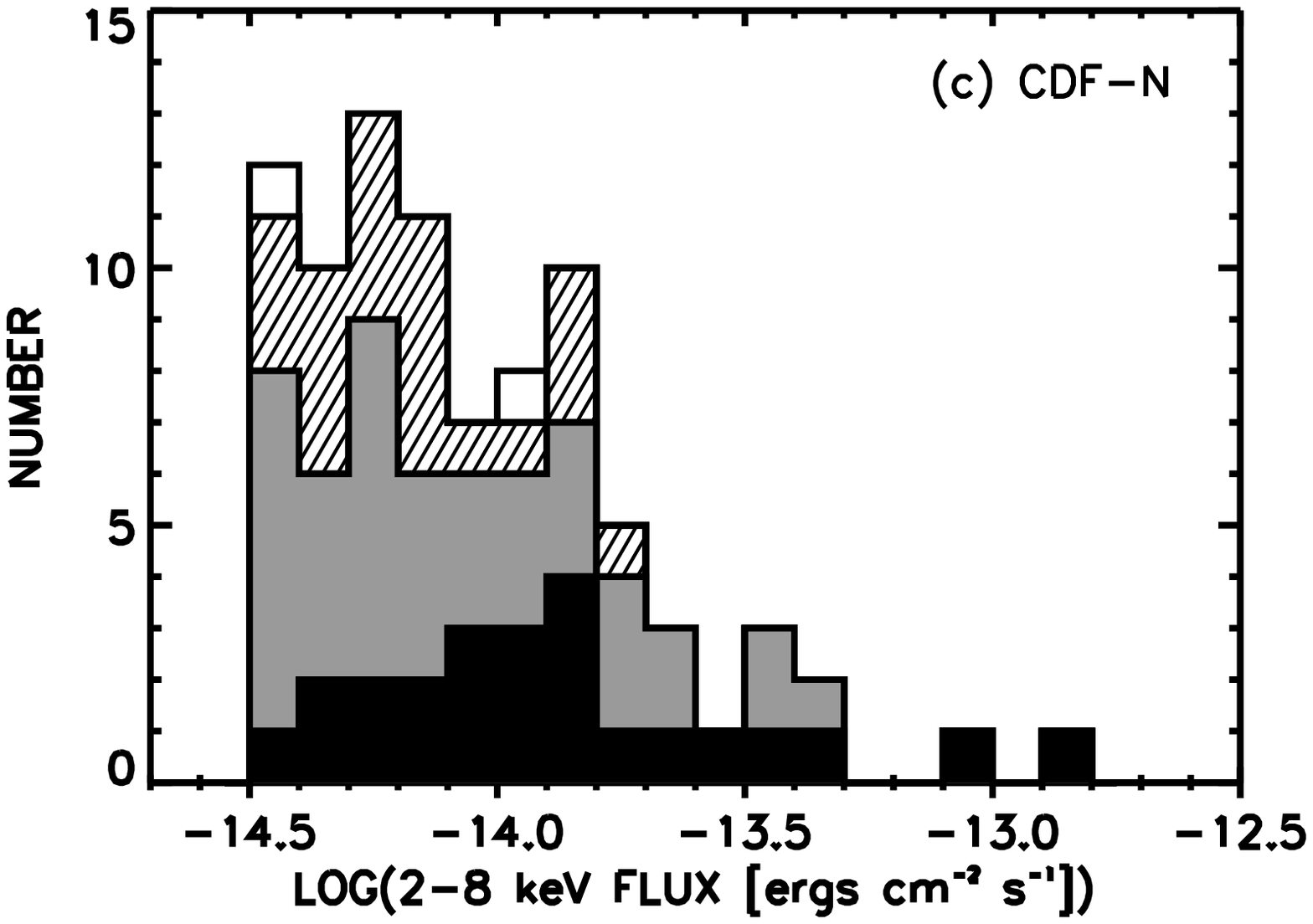}{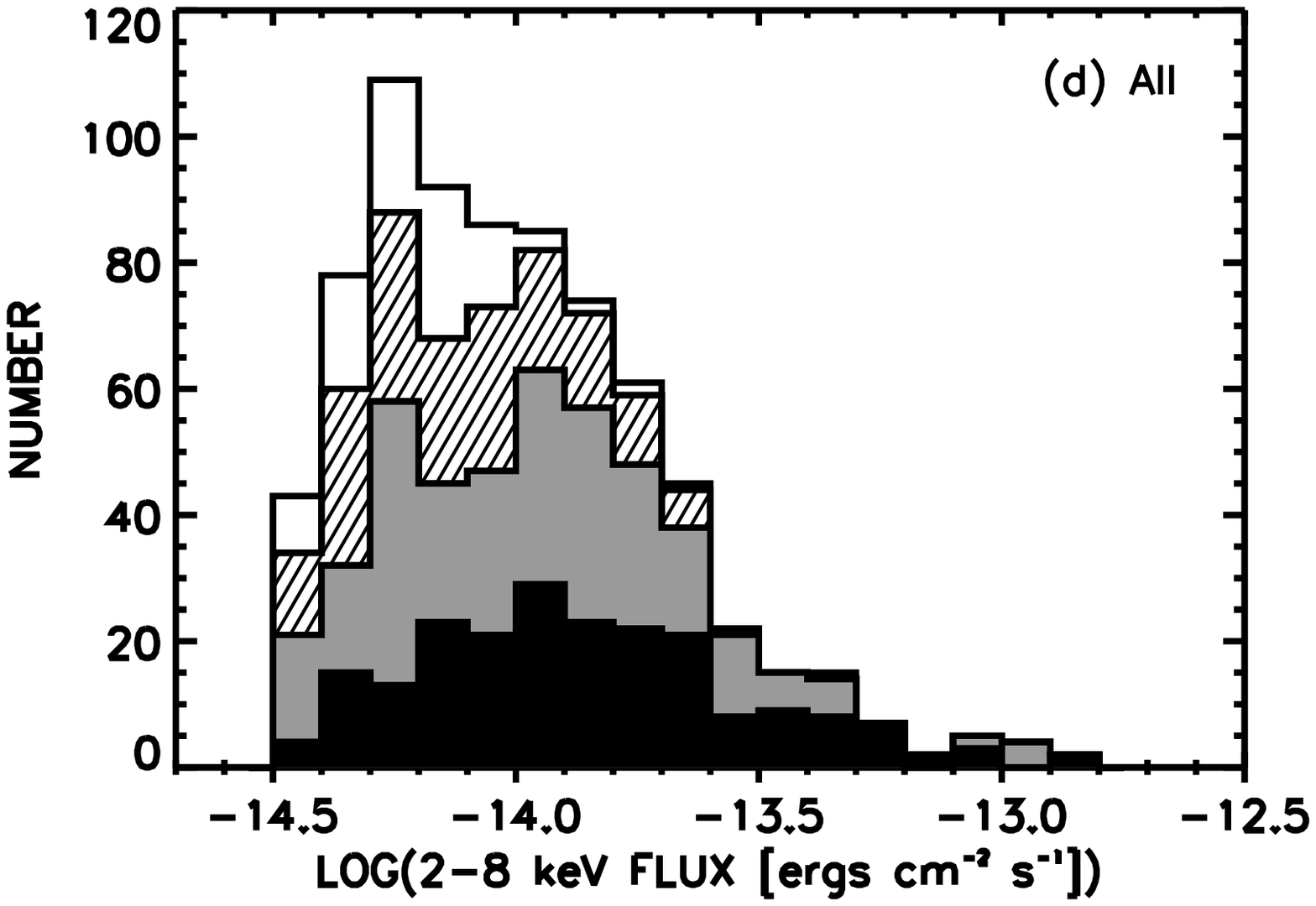}
\caption{$2-8$~keV flux distributions 
for our (a) CLANS, (b) CLASXS, (c) CDF-N, and (d) total
$2-8$~keV sample (black, spectroscopically identified 
BLAGNs; shaded, spectroscopically identified 
non-BLAGNs; hatched, spectroscopically observed 
but unidentified sources; open, spectroscopically unobserved 
sources).}
\label{spect}
\end{figure}

\citet{trouille08} present the details of our spectroscopic observations 
of the X-ray sources in the CLANS, CLASXS, and CDF-N fields, including
how the observations were made, the data reduction process, and the latest 
redshift catalogs. 

We provide a catalog of updated values for the CLANS field in Table
\ref{clansupdate}. We ordered the sources by increasing right
ascension and labeled each with the same source number (col [1]) as in
\citet{trouille08}. CLANS \#761 through \#789 were inadvertently not
included in the original \citet{trouille08} CLANS catalog, and so are
included here. CLANS \#80, \#150, and \#404 in the \citet{trouille08}
catalog had incorrect $2-8~\rm keV$ fluxes as a result of their
anomolously high hardness ratio values. Here we list their corrected fluxes
determined using the HEASARC WebPIMMS tool and assuming
$\Gamma=1.4$. The remaining sources are included as a result of our
spectroscopic observations with DEIMOS on Keck II in Spring 2009. 

Columns (2) and (3) give the right ascension and declination
coordinates. Columns (4) and (5) list the net counts in the $0.5-2~\rm
keV$ and $2-8~\rm keV$ bands. Columns (6), (7), and (8) provide the
X-ray fluxes in the $0.5-2~\rm keV$, $2-8~\rm keV$, and $0.5-8~\rm
keV$ bands, respectively, in units of
$10^{-15}$~erg~cm$^{-2}$~s$^{-1}$. The errors quoted are the
1~$\sigma$ Poisson errors, using the approximations from
\citet{gehrels86}. The flux errors do not include the uncertainty in the flux
conversion factor; however, the errors are generally dominated by the Poisson
errors. Column (9) gives the spectroscopic redshifts and column (10)
gives the optical spectral classifications. 

For $f_{2-8~{\rm keV}}>1.4\times10^{-14}$~erg~cm$^{-2}$~s$^{-1}$ (the 
break flux in the $2-8$~keV number counts; see \citealt{trouille08}), 
we have spectroscopic redshifts for 177 of the 208 sources in our 
$2-8$~keV sample. We classified our spectroscopically identified
sources into the four optical spectral types given in the Introduction.  
We list the number of sources of each spectral type by field in
Table~\ref{hclass table}. The percentages 
refer to the percent of identified sources having that spectral type.

In Figure~\ref{spect} we show the flux distributions by field
and all together for the BLAGNs (black), the non-BLAGNs (shaded), 
the spectroscopically observed but unidentified sources (hatched), 
and the spectroscopically unobserved sources (open) in our 
$2-8~\rm keV$ sample.
The higher spectroscopic completeness at bright X-ray fluxes is 
partly due to the fact that at these fluxes the sample is dominated 
by BLAGNs, which are straightforward to identify. In addition, at
fainter X-ray fluxes the
sources tend to be optically fainter, making the redshift
identifications at these fluxes more difficult.
In particular, the
intermediate-flux, optically normal galaxies at $z\sim2$ are the 
most difficult to identify spectroscopically.

\begin{figure}[!ht]
\epsscale{1.2}
\plotone{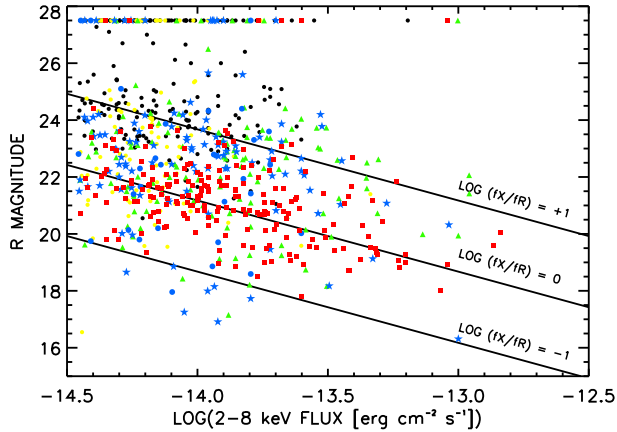}
\caption{$R$ magnitude vs.~$2-8$~keV flux for our 
$2-8$~keV sample (red squares, BLAGNs; green triangles, 
high-excitation sources; blue circles, absorbers; blue stars, 
star formers; black circles, spectroscopically observed 
but unidentified sources; yellow circles, spectroscopically 
unobserved sources.
Solid lines show where the fluxes are
equal [i.e., $\log(f_X)/f(R)=0$] and the boundaries of the region 
where AGNs typically reside [$\log(f_X/f_R)=\pm 1$].
Magnitudes brighter than $R\sim 20$ suffer from saturation problems
and are likely to be underestimated.  Sources undetected at the 
$2~\sigma$ limits of the images are plotted at $R=27.5$.
\label{Rvsfx}}
\end{figure}

\begin{figure}[!ht]
\epsscale{1.2}
\plotone{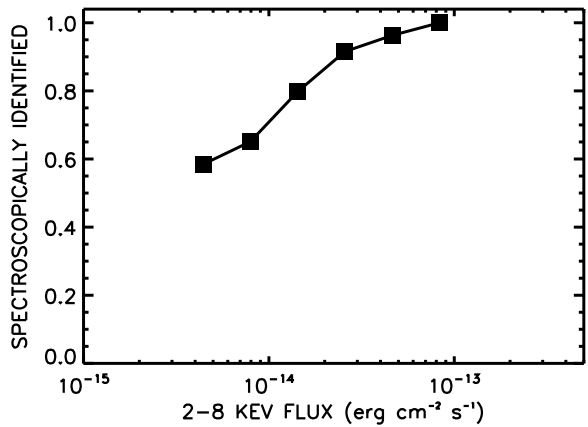}
\label{hfrac}
\caption{Fraction of spectroscopically observed sources in our $2-8$~keV 
sample that are spectroscopically identified
in bins of $2-8$~keV flux. Only bins which 
contain five or more sources are plotted.  Due to the
complicated observing program for the CDF-N, all sources
in that field are treated as spectroscopically observed, though we
note that there are only two sources that have not been observed (see
Figure \ref{spect}).}
\end{figure}

In Figure~\ref{Rvsfx} we show $R$ magnitude versus 
X-ray flux for our $2-8$~keV sample.  AGNs typically lie in 
the region defined by the loci $\log(f_X/f_R)=\pm 1$
(e.g., \citealt{maccacaro88}; \citealt{schmidt98}; 
\citealt{hornschemeier01}; \citealt{alexander02}; 
\citealt{bauer02}; \citealt{barger03}).
As expected, the majority of our spectroscopically observed but
unidentified sources (black circles) have faint optical 
magnitudes ($>70\%$ have $R>24$). 
In obtaining our spectroscopy, we selected the sources without 
any regard to their optical magnitudes in order to avoid additional 
selection effects.  This is evident from the figure, where 
the spectroscopically unobserved sources (yellow circles) cover 
the entire range of optical magnitudes.  We will not show these 
sources in our subsequent figures nor use them in our subsequent
analysis, as they can be considered a random sample of the population
and will not affect our results.  
Thus, in Figure~\ref{hfrac} we show the fraction of spectroscopically
observed sources in our $2-8$~keV sample that are spectroscopically 
identified versus X-ray flux.

In \citet{trouille08} we extended the redshift information for the
three fields by determining photometric redshifts. 
The CLANS field has eight bands of coverage ($g'$, $r'$, $i'$, $z'$,
$J$, $H$, $K$, $3.6~\mu$m), the CLASXS field has 11 bands of coverage
($u$, $B$, $g'$, $V$, $R$, $i'$, $z'$, $J$, $H$, $K$, $3.6~\mu$m), and
the CDF-N field has 10 bands of coverage ($U$, $B$, $V$, $R$, $I$,
$z'$, $J$, $H$, $K_s$, $3.6~\mu$m). 
We used the template-fitting method described in 
\citet[and references therein]{wang06}. We built our training 
set of spectral energy distributions (SEDs) using the spectroscopically 
identified sources in our sample. We then determined
the photometric redshifts by finding the best fit (via least-squares 
minimization) between our individual source SEDs (using the
optical through infrared data) and these templates. 
We include the number of photometric redshifts for the spectroscopically
observed but unidentified sources obtained in this way as an entry 
in Table~\ref{hclass table}.

\begin{figure}[!ht]
\epsscale{1.2}
\plotone{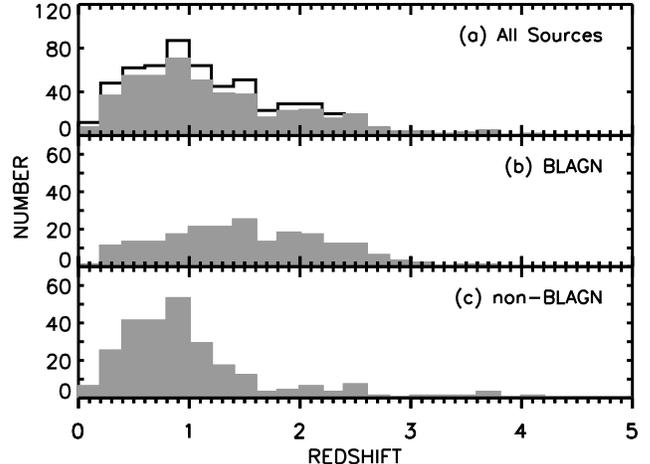}
\caption{Redshift distributions for our spectroscopically observed
$2-8$~keV sample by optical spectral type.  The shaded area shows 
(a) all spectroscopic redshifts, (b) BLAGNs, and 
(c) non-BLAGNs.  The solid line in (a) shows the spectroscopic 
plus photometric redshift distribution.}
\label{optz}
\end{figure}

In Figure~\ref{optz} we show the redshift distributions for
our spectroscopically observed $2-8$~keV sample by optical spectral 
type (all, BLAGN, non-BLAGN).  The shaded areas show the 
spectroscopically identified distributions, while the solid line 
in (a) shows the spectroscopically plus photometrically identified 
distribution.  The non-BLAGN redshift distribution is
strongly peaked at $z\sim 0.9$.

\section{Comparison of Optical Classification with X-ray Classification} 
\label{gamma}

\subsection{\Geff~Distributions}
 \label{gdistribution}

\begin{figure}[!ht]
\epsscale{1.2}
\plotone{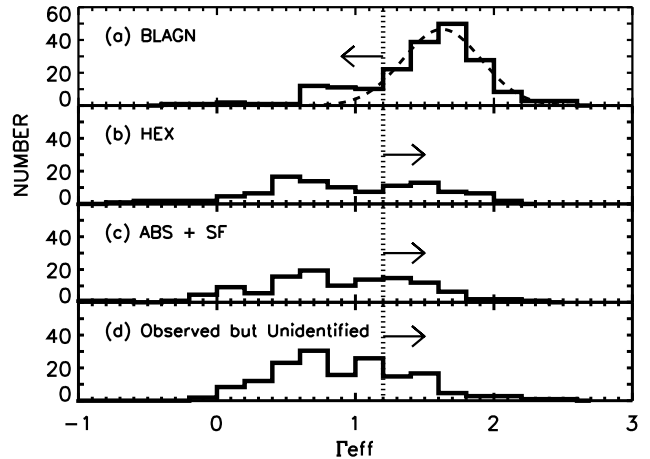}
\caption{\Geff~distributions for the BLAGNs, the
  high-excitation sources, the absorbers plus star formers, and
  the unidentified sources in our spectroscopically observed
  $2-8$~keV sample. The dashed curve shows the gaussian fit to 
  the \Geff~distribution for the BLAGNs
  ($\langle\Gamma_{\rm eff}\rangle=1.63$ and $\sigma=0.28$). 
  The vertical dotted line shows 
  $\Gamma_{\rm eff}=1.2$. The arrows indicate the locations of the 
  outlier sources discussed in Section~\ref{gamma}.}
\label{disp}
\end{figure}

In Figure~\ref{disp} we show the \Geff~distributions for
the spectroscopically observed sources in our
$2-8$~keV sample by optical spectral type 
(BLAGN, HEX, ABS+SF, unidentified).
For the CLANS sources we used the \Geff~values given in 
\citet{trouille08}, and for the CDF-N sources we used the 
\Geff~values given in \citet{alexander03}. 
To determine the \Geff~values for the CLASXS sources we followed 
the method used by \citet{trouille08}. In short, from the
HRs given in \citet{yang04} and using XSPEC, we determined the 
HR-to-\Geff~conversion by assuming a
single power law spectrum with Galactic absorption. 

We see from Figure~\ref{disp} that there is
substantial overlap in the \Geff~distributions for the
BLAGNs and the non-BLAGNs.  We also note that the spectroscopically 
unidentified sources have a \Geff~distribution which is very 
similar to the HEX and ABS+SF sources.

The dashed curve in Figure~\ref{disp}(a) shows the gaussian fit 
to the \Geff~distribution for the BLAGNs.
We find that the BLAGNs are clustered around $\langle \Gamma_{\rm eff}
\rangle = 1.63$ 
with a dispersion $\sigma=0.28$. \citet{mushotzky84} was the first to 
note this narrow range in photon indices. The vertical line at 
$\Gamma_{\rm eff}=1.2$ indicates the value above which 80\%
of the BLAGNs reside. As noted in the Introduction, 
$\Gamma_{\rm eff}=1.2$ is approximately equivalent to HR $= -0.2$, 
which is the value \citet{szokoly04} chose to use
to distinguish between X-ray type~1 (unabsorbed) sources and 
X-ray type~2 (absorbed) sources.

\subsubsection{Optically Unobscured but X-ray Absorbed}
\label{outlier1}

As is well known (see references in the Introduction), despite 
the generally good agreement between the two classification schemes 
regarding which sources are unabsorbed, the agreement is not 100\%.
There are BLAGNs whose X-ray spectral properties suggest substantial 
absorption.  In our spectroscopically observed $2-8$~keV sample, 
we find that 20\%$\pm 3$\% of the 207 BLAGNs have \Geff~$<1.2$.
The leftward pointing arrow in Figure~\ref{disp}(a) calls
attention to these BLAGNs with low values of \Geff,
all of which have at least one emission line with 
FWHM~$>2000$~\kms. The optical spectra of these sources
do not appear to share any common characteristics, nor
do they appear any different than the spectra for our BLAGNs 
with $\Gamma_{\rm eff} \ge 1.2$.  
Only one of these sources (CLASXS~\#174) has less than 10 counts 
at $0.5-8$~keV and may suffer from contamination by 
uncleared afterglow events 
(see http://asc.harvard.edu/ciao/why/afterglow.html). 

\citet{kuraszkiewicz09a,kuraszkiewicz09b} studied a
sample of red AGNs that they selected from the Two Micron 
All Sky Survey (2MASS) on the basis of their red $J-K_s$ 
colors ($>2$~mag) and then followed up with {\em Chandra\/}.
85\% of their $J-K>2$ sample show broad-lines, 
and the remainder are narrow-line AGNs. 
Through detailed modeling they found that the shape of the 
SEDs for these sources was generally consistent with modest 
absorption by gas (in X-rays) and dust (in the optical/NIR). Using
principal component analysis, they argue that the Eddington ratio (and
not absorption by the circumnuclear material or the host galaxy) is
the dominant factor in determining the shape of the SED. 

\begin{figure}
\epsscale{1.1}
\plotone{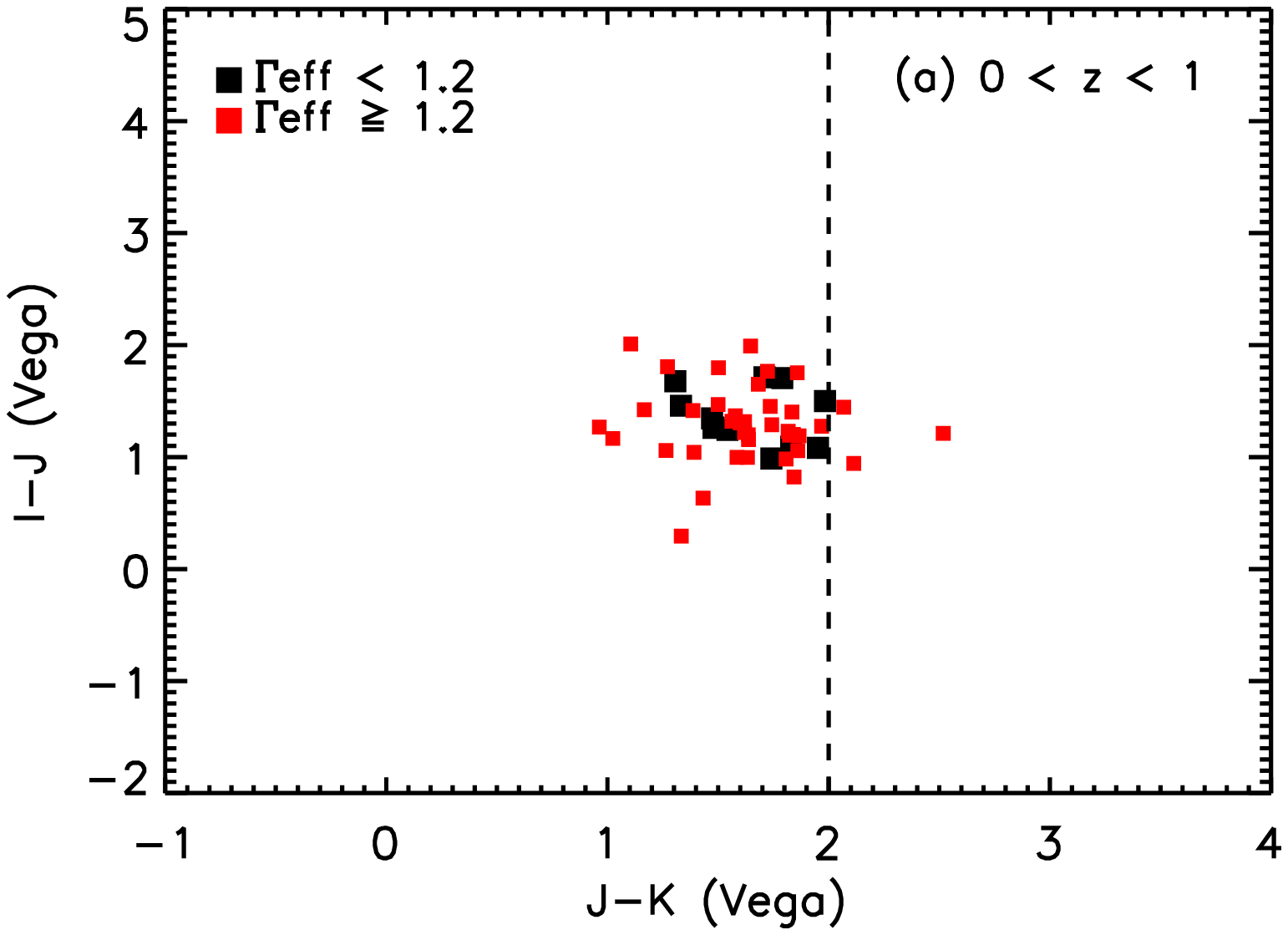}
\plotone{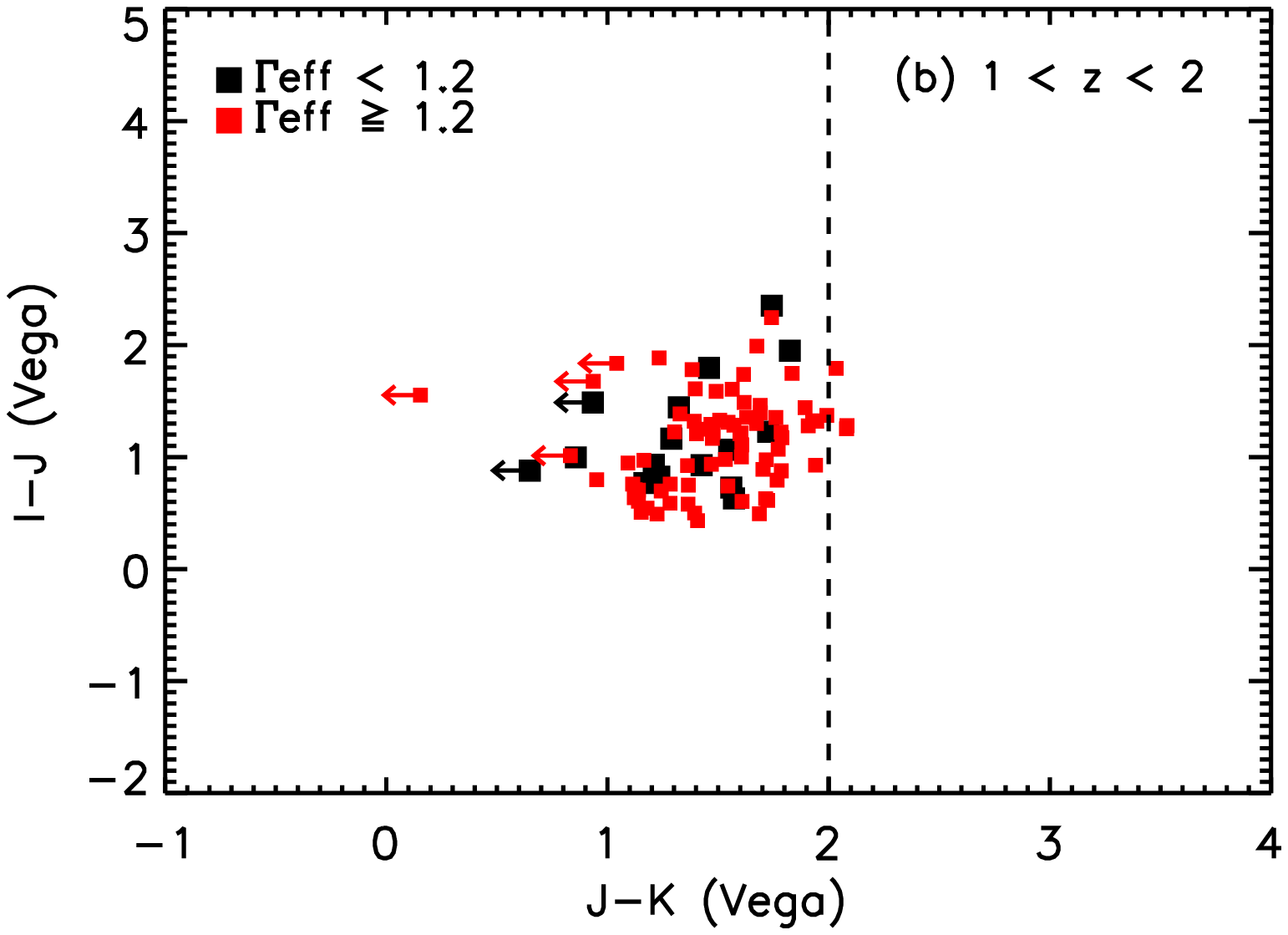}
\plotone{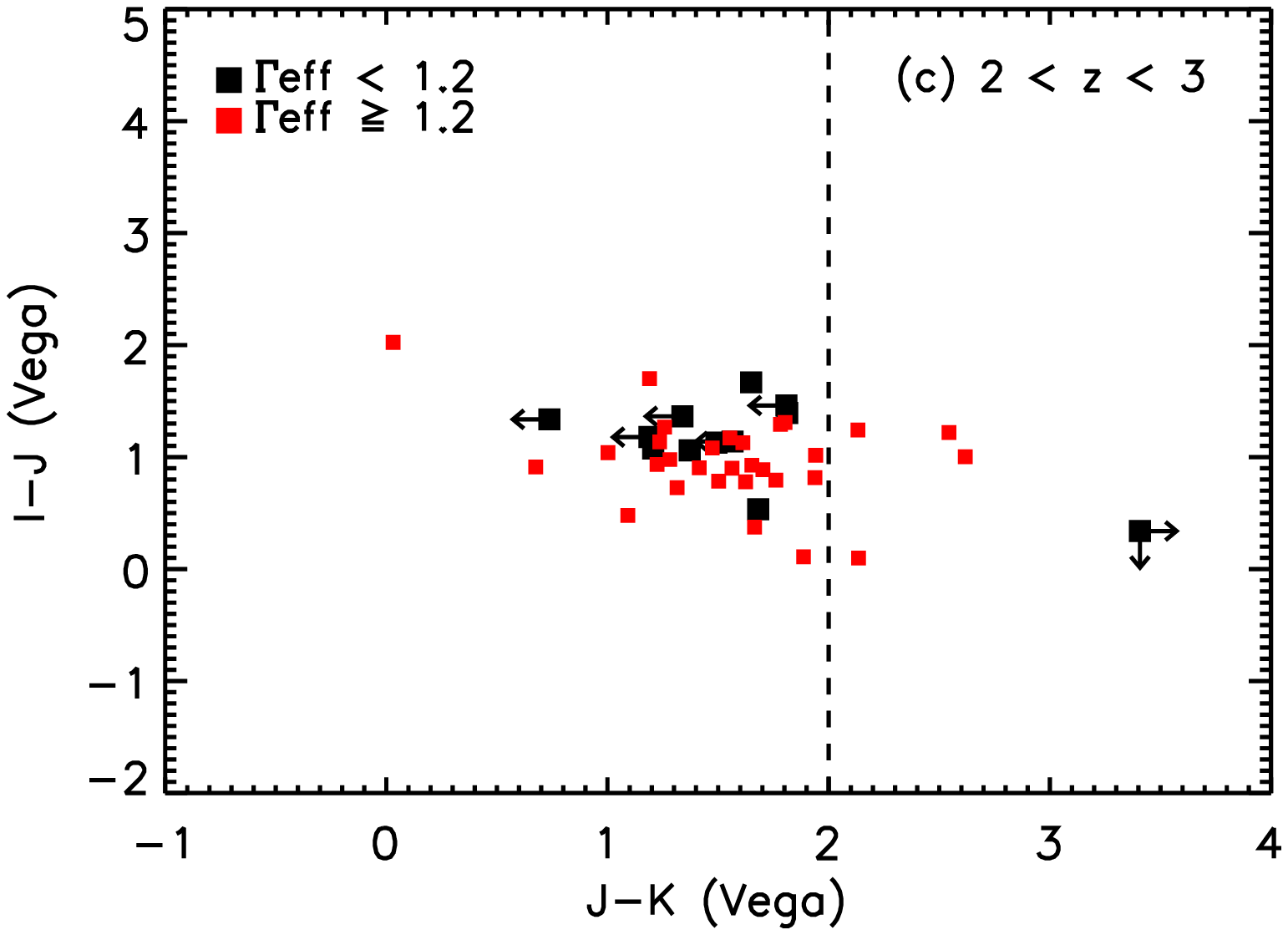}
\caption{$I-J$ vs. $J-K$ for the BLAGNs in our spectroscopically
observed $2-8$~keV sample in the redshift intervals 
(a) $0<z<1$, (b) $1<z<2$, and (c) $2<z<3$. BLAGNs with
$\Gamma\ge 1.2$ are shown in red, while BLAGNs with
$\Gamma<1.2$ are shown in black. 
Sources undetected in two or more of the three bands are not 
shown.  Sources with leftward
pointing arrows were not detected above the limiting 
magnitude in the $K$ band of their respective field. 
The source in (c) with a
rightward pointing arrow and a downward pointing arrow
was not detected above the limiting magnitude in the $J$ 
band of the CLANS field (CLANS \#301). 
\label{IJvsJK}}
\end{figure}

We can use our color information to look for color differences between
our X-ray absorbed and X-ray unabsorbed BLAGNs to test for extinction
associated with low \Geff~values (assumming the low
\Geff~value is due to absorption by gas along the line-of-sight). In
Figure~\ref{IJvsJK} we show $I-J$
versus $J-K$ for the BLAGNs in our spectroscopically observed $2-8$~keV
sample. To avoid $K-$correction effects in the comparison, we divide
our sample by redshift. In each redshift interval we see no
significant differences in the color distributions of the \Geff~$<1.2$ and
\Geff~$\ge 1.2$ BLAGNs. This provides support for the
suggestion that there is not a one-to-one
correspondence between \Geff~and extinction (\citealt{cowie09};
see also \citealt{maiolino01b} and \citealt{kuraszkiewicz09b}).

\subsubsection{Optically Obscured but X-ray Unabsorbed}
\label{outlier2}

As is also well known (see references in the Introduction),
there are a number of sources whose optical spectra do not show broad 
lines, suggesting obscuration, while their X-ray spectral properties 
suggest little absorption.  In our spectroscopically observed $2-8$~keV 
sample, 33\%$\pm 4$\% of the 255 non-BLAGNs have $\Gamma_{\rm eff}\ge 1.2$.
The rightward pointing arrows in Figures~\ref{disp}(b), (c), and (d)
call attention to these non-BLAGNs with high values of \Geff. 
The optical spectra for these sources do not appear to
share any common characteristics, nor do they
appear any different than the spectra for our non-BLAGNs with 
$\Gamma_{\rm eff}<1.2$. All of these sources have more than 10 counts
at $0.5-8$~keV. 

We may look at the most extreme examples of conflicting
optical and X-ray information by considering the 22 non-BLAGNs
with \Geff~$\ge1.7$. Considering in detail the optical spectra of the
individual non-BLAGNs in our $2-8$~keV sample having $\Gamma_{\rm
  eff} \ge 1.7$, we find that all have spectra covering at least one of
the potentially broadened emission lines (i.e., H$\alpha$, H$\beta$,
H$\gamma$, MgII, Ly$\alpha$, and CIV). All but two of the HEX
sources exhibit FWHM~$<1500$~\kms~(CDF-N~\#301 and CDF-N~\#331 exhibit
$1500<$~FWHM~$<1900$~\kms). Moreover, all of the optically
spectroscopically identified absorbers can be
confidently classified as non-BLAGNs, as can all but one of the star 
formers (CLANS \#71 lacks sufficient signal-to-noise at the wavelengths
of the potentially broadened emission lines covered by its spectrum).

\subsection{\Geff~versus X-ray Luminosity}

In Section~\ref{gdistribution} we found that while $80$\% $\pm$ 6\% of 
the BLAGNs in our sample have \Geff~$\ge1.2$, 33\% $\pm$ 4\% of our non-BLAGNs
also have \Geff~$\ge1.2$. We now expand our spectroscopically observed
$2-8$~keV sample
to include \emph{ASCA} large-area, bright-flux data (hereafter, 
we refer to this larger sample as
our ``extended spectroscopically observed $2-8$~keV sample''). 
These additional data are needed to have a sufficiently large 
volume at the high-luminosity, low-redshift end.

We use the optical spectral classifications, redshifts, X-ray
luminosities, and \Geff~values from \nocite{akiyama00,akiyama03}
Akiyama et al.~(2003; 75 optically identified AMSSn AGNs)
and Akiyama et al.\ (2000; 30 optically identified 
ALSS AGNs). We converted their $2-10$~keV
luminosities into rest-frame $2-8~\rm keV$ luminosities using
\begin{equation}
L_{2-8~\rm keV} = \frac{\int^{8~\rm keV}_{2~\rm keV}\nu^{-\Gamma_{\rm eff}}~ 
d\nu}{\int^{10~\rm keV}_{2~\rm keV}\nu^{-\Gamma_{\rm eff}} ~d\nu}  
\times L_{2-10~\rm keV}
\end{equation}
and the \Geff~for each source. $L_{2-8~\rm keV}$ is on average 
$\approx90$\% of $L_{2-10~\rm keV}$. 
We calculated the rest-frame $2-8$~keV luminosities for our three OPTX
fields using $L_X=f \times 4 \pi d_L^2 \times K-$correction.
Assuming an intrinsic $\Gamma=1.8$, we used
\begin{eqnarray*}
{\rm for} \, z<3,\,\, K{\rm -corr} &=& (1+z)^{-0.2}\, {\rm and}\, f=f_{2-8~\rm keV} \,, \\
{\rm for} \, z \ge 3,\,\, K{\rm -corr} &=& \Biggl[\frac{1}{4}(1+z)\Biggr]^{-0.2}\, {\rm and}\, f=f_{0.5-2~\rm keV} \,.
\end{eqnarray*}
The $\frac{1}{4}$ factor in the $z\ge 3$ $K-$correction is a result of
normalizing so that there is no $K-$correction when $z=3$, at which point
observed-frame $0.5-2$~keV corresponds exactly to rest-frame
$2-8$~keV.
 
\begin{figure}[!ht]
\epsscale{1.1}
\plotone{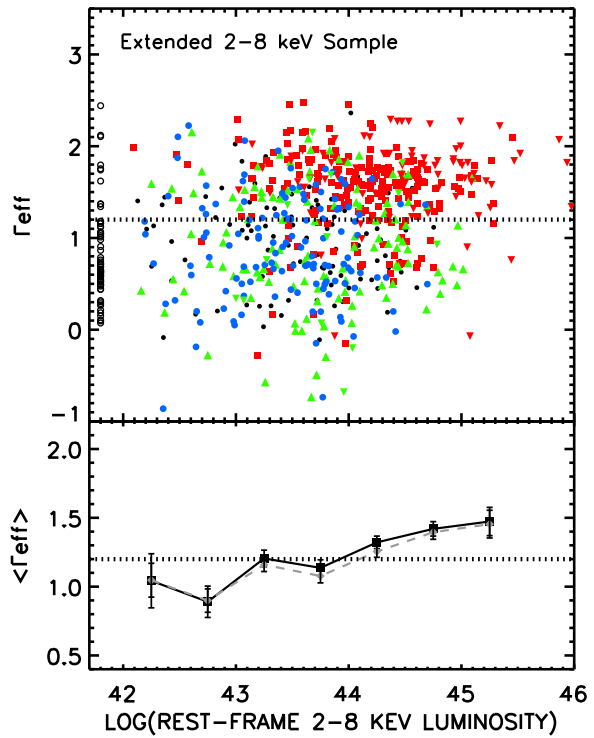}
\caption{(Upper panel) \Geff~vs.~rest-frame $2-8$~keV luminosity
  for the sources in our extended spectroscopically observed $2-8$~keV 
  sample (red squares, BLAGNs; 
  red inverted triangles, {\em ASCA} BLAGNs;
  green triangles, high-excitation sources; 
  green inverted triangles, {\em ASCA} non-BLAGNs;
  blue circles, absorbers and
  star-formers; black solid circles, spectroscopically
  observed but unidentified sources with photometric redshifts;
  black open circles plotted at a nominal luminosity, spectroscopically
  observed but unidentified sources without photometric redshifts). 
  (Lower panel) $\langle$\Geff$\rangle$~in luminosity bins
  vs.~rest-frame $2-8$~keV luminosity.  The spectroscopic sample
  is denoted by the black symbols and solid lines, while the 
  spectroscopic plus photometric sample is denoted by 
  the gray symbols and dashed lines.  Only bins with at least 
  six sources are plotted. 
  The error bars show $\frac{1~\sigma}{\sqrt{N}}$. 
  The dotted horizontal lines in both panels show $\Gamma_{\rm eff}=1.2$.}    
\label{GLx}
\end{figure}

In Figure~\ref{GLx} (upper panel) we plot \Geff~versus rest-frame 
$2-8$~keV luminosity for our extended spectroscopically observed
$2-8$~keV sample, and in Figure~\ref{GLx} (lower panel) we plot
$\langle$\Geff$\rangle$ versus rest-frame $2-8$~keV luminosity.  
According to \citet{hasinger05}'s mixed classification scheme described 
in the Introduction, any source above the dashed horizontal line 
($\Gamma_{\rm eff}=1.2$)
would be considered unabsorbed, and any source below 
(other than optically classified BLAGNs) would be considered
absorbed. (Of course, to be able to exclude the BLAGNs that lie 
below the line from the absorbed category---or to be able to 
include them in the unabsorbed category---requires high 
optical spectroscopic completeness.) 
We can see that their method picks up most of its `new' (above 
the line) unabsorbed sources (as compared 
with what would be found from a pure optical classification scheme) 
at lower X-ray luminosities.  However, it also picks up a large number 
of HEX sources at higher X-ray luminosities.  Since we do not yet have
a good understanding of how the X-ray and optical classification schemes 
relate to the obscuration of the central engine, mixing the two
classification schemes in this manner can only complicate the 
interpretation, and we strongly advise against it.

From Figure~\ref{GLx} (upper panel) we see the well-known
luminosity dependence of the optical spectral types,
with the BLAGNs dominating the numbers at high X-ray luminosities
\citep{lawrence82,steffen03,
ueda03,barger05,lafranca05,simpson05,akylas06,beckmann06,sazonov07,
dellaceca08,hasinger08,silverman08,winter09,yencho09}. 
From Figure~\ref{GLx} (lower panel) we see there is also
a luminosity dependence of the effective X-ray photon index, 
with $\langle$\Geff$\rangle$ rising from $\sim1$ at 
$L_X=10^{42}$~\ergss~to $\sim1.5$ at $L_X=10^{45}$~\ergss. 

\begin{figure}
\epsscale{1.1}
\plotone{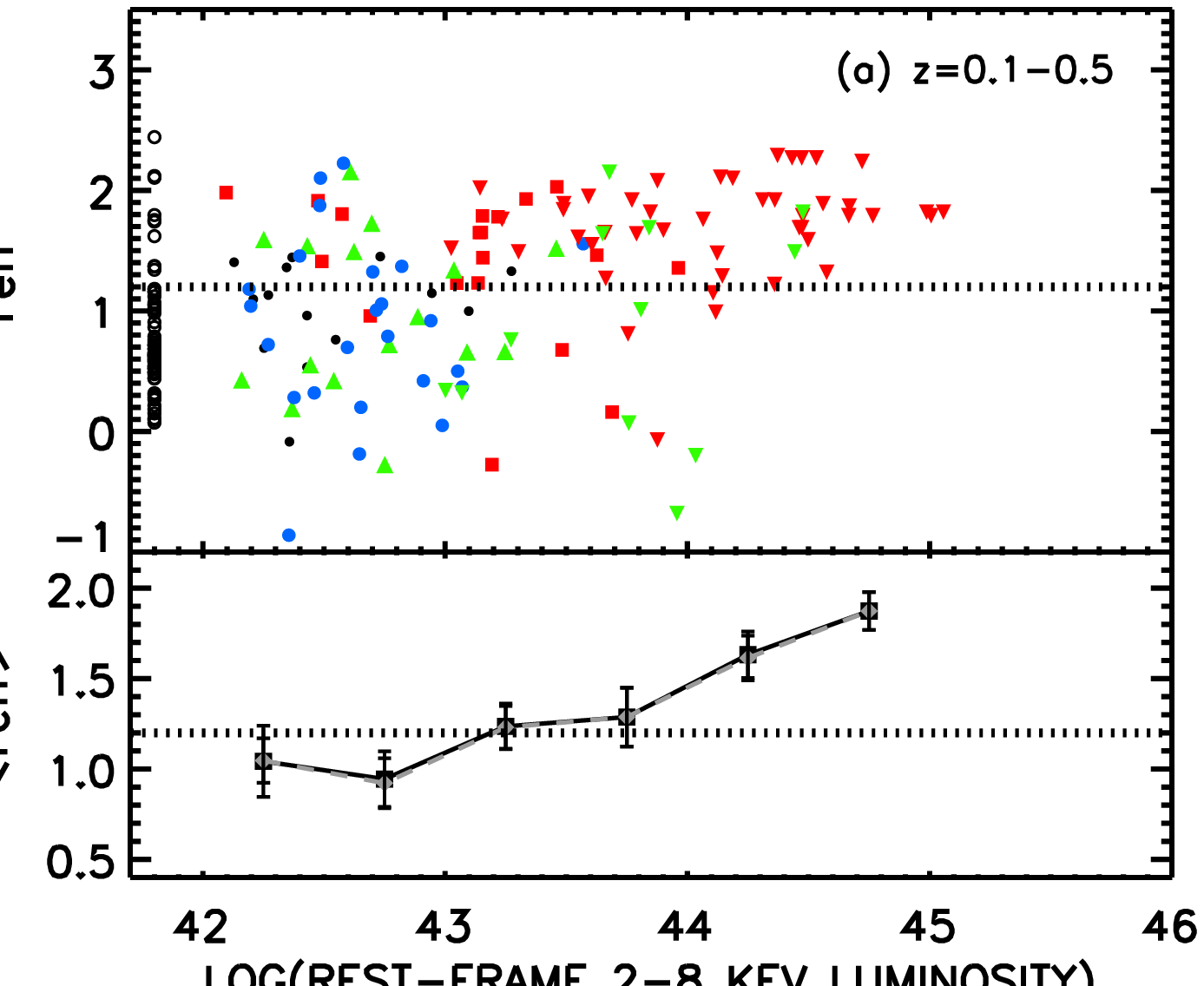}
\plotone{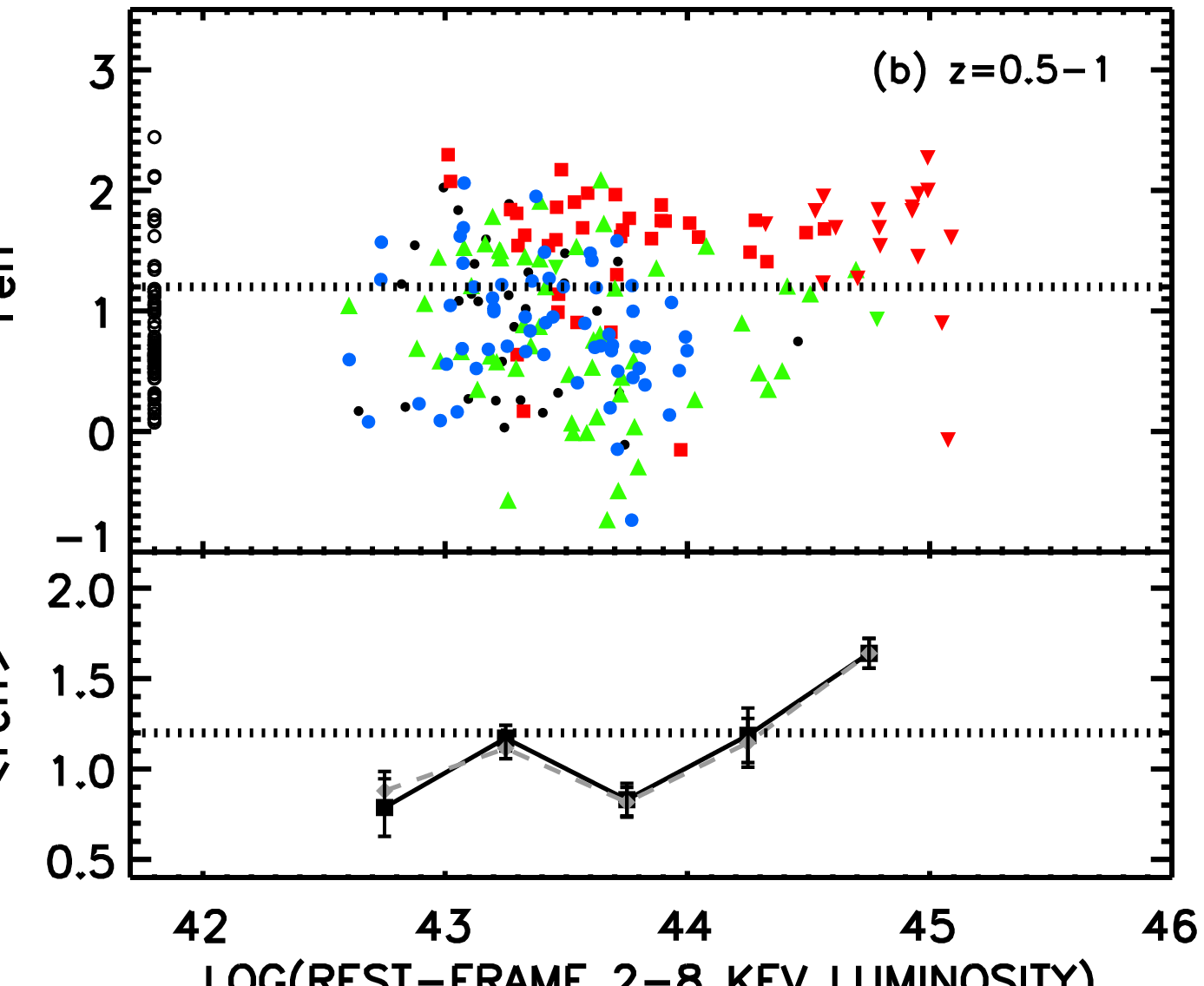}
\plotone{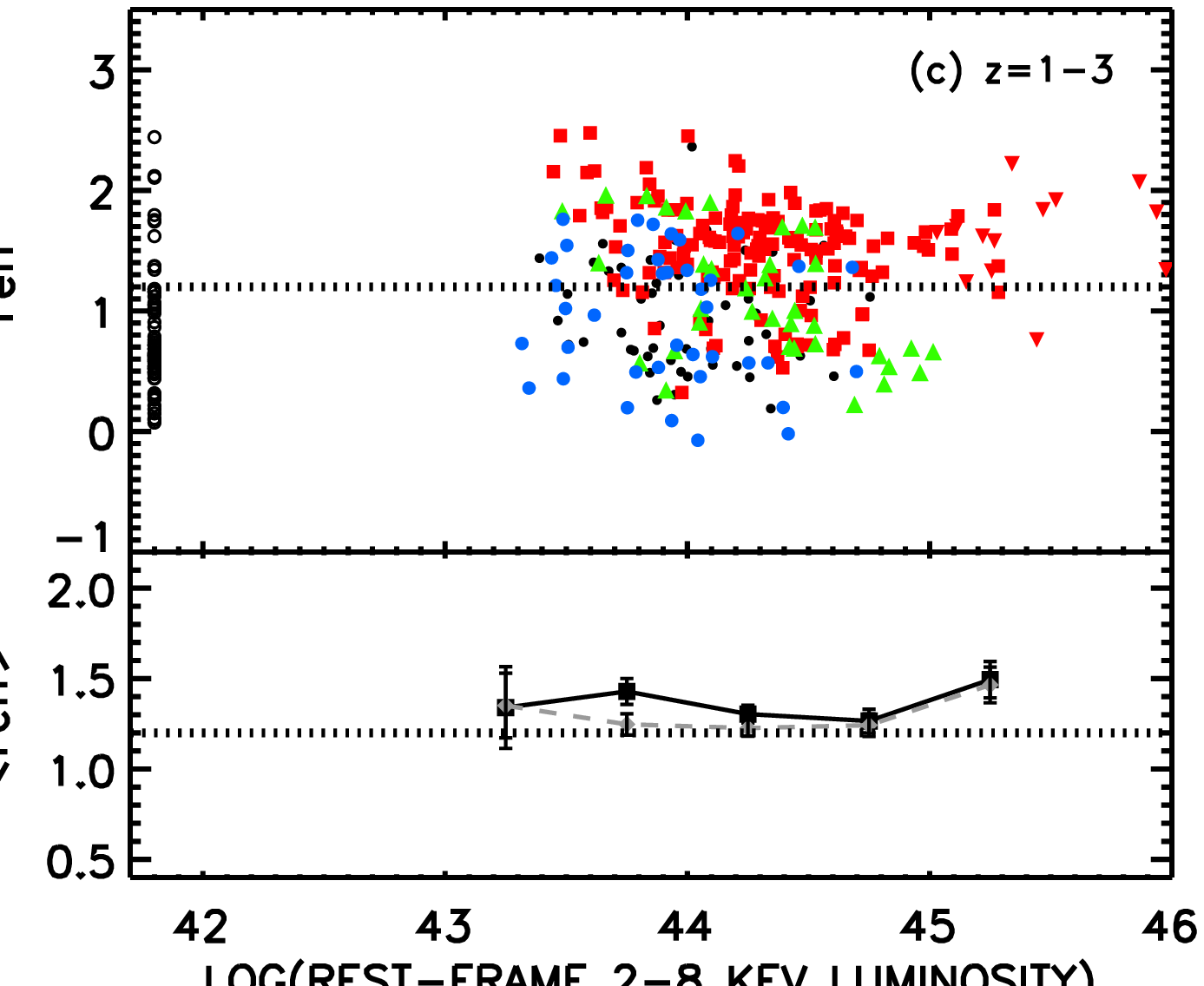}
\caption{(Upper panels) \Geff~vs.~rest-frame $2-8$~keV
  luminosity for the sources in our extended spectroscopically
  observed $2-8$~keV sample with 
  (a) $z=0.1-0.5$, (b) $z=0.5-1$, and (c) $z=1-3$. 
  (Lower panels) $\langle\Gamma_{\rm eff}\rangle$ in luminosity bins 
  vs.~rest-frame $2-8$~keV luminosity.  Only bins with at least six 
  sources are plotted. The error bars show $\frac{1~\sigma}{\sqrt{N}}$. 
  In both sets of panels the symbols are the same as in Figure~\ref{GLx}.
\label{gammaLx}
}
\end{figure}

In Figure~\ref{gammaLx} we divide Figure~\ref{GLx}
into three redshift intervals: (a) $z=0.1-0.5$,
(b) $z=0.5-0.1$, and (c) $z=1-3$.  
Although the distributions in the upper panels of the
figure look the same for each redshift
interval, we can see that the transition luminosity from X-ray 
soft ($\Gamma_{\rm eff}\ge1.2$)
dominated to X-ray hard dominated (or from BLAGN dominated to
non-BLAGN dominated) is shifting to higher luminosities 
with increasing redshift interval.  This redshift dependence
of the transition luminosity was first noted by
Barger et al.\ (2005; see their Figure~19)\nocite{barger05} from the
luminosity functions.

In the lower panels of Figure~\ref{gammaLx} we see that
for the two lower redshift intervals there is a dominance 
of the X-ray soft sources at higher luminosities and a
decreasing influence of these sources at lower luminosities,
as previously noted from Figure~\ref{GLx}. 
However, in the $z=1-3$ interval we instead see a
continued dominance of the soft sources at lower luminosities.
This is in part due to the fact that at higher redshifts the 
$2-8$~keV band samples higher energies, and these higher energies
are less affected by obscuring material (see \citealt{kim07b} for
modeling of this effect).  Thus,
a higher percentage of $z>1$ non-BLAGNs are X-ray soft 
(have high \Geff) compared with those at $z<1$, as
\citet{szokoly04} cautioned when they were defining their
X-ray classification scheme (see Introduction).  This emphasizes
the danger in using the highly redshift-dependent hardness 
ratio for classifying sources as absorbed or unabsorbed, as
is done in both the pure X-ray classification and mixed 
classification schemes.

In Table~\ref{byz} we give the percentages of high-excitation
sources, star formers, and absorbers with $z<1$ and $1<z<3$ in
our extended spectroscopically observed $2-8$~keV sample with 
\Geff$\ge1.2$, \Geff$\ge1.5$, and \Geff$\ge1.7$.  
The non-BLAGNs in the $1<z<3$ redshift interval exhibit a higher 
percentage of X-ray soft sources than the $z<1$ redshift interval 
(the exception being the absorbers, which suffer from small numbers 
in the $z=1-3$ interval). 

\begin{table}
\begin{small}
\centering
\caption{Non-BLAGNs with Soft Photon Indices in Our
Extended Spectroscopically Observed $2-8$~keV Sample by Redshift
Interval}
\label{byz}
\begin{tabular}{l c c c }
\tableline\tableline
\Geff & HEX\tablenotemark{a} & SF\tablenotemark{a} & ABS\tablenotemark{a}\\
\tableline
\multicolumn{4}{c}{$z<1$}   \\
\tableline
All    & 74        & 60        & 30\\
$\ge1.2$ & 27 (36\%) & 16 (26\%) & 7 (23\%)   \\
$\ge1.5$ & 15 (20\%) & 6 (10\%) & 4 (13\%)   \\
$\ge1.7$ & 7 (10\%) & 3 (5\%) & 2 (7\%)   \\
\tableline
\multicolumn{4}{c}{$1<z<3$} \\
\tableline
All    & 37        & 41        & 3\\
$\ge1.2$ & 15 (41\%) & 18 (44\%) & 0 (0\%)   \\
$\ge1.5$ & 9 (24\%) & 8 (20\%)   & 0 (0\%)   \\
$\ge1.7$ & 7 (19\%) & 3  (7\%) & 0 (0\%)   \\
\tableline
\end{tabular}
\end{small}
\footnotesize
\tablenotetext{a}{The percentages refer to the percent of 
sources of the given optical spectral type with the specified 
values for \Geff.}
\end{table}

\section{Discussion}
\label{discussion}

Using our large spectroscopically observed $2-8$~keV sample 
with $>80$\% spectroscopic 
completeness for $f_{2-8~\rm keV}>10^{-14}$~erg~cm$^{-2}$~s$^{-1}$
and $>60$\% spectroscopic completeness down to our chosen flux 
limit of $f_{2-8~\rm keV}=3.5\times 10^{-15}$~erg~cm$^{-2}$~s$^{-1}$,
we have confirmed that there is considerable overlap of the X-ray 
spectral properties for different optical spectral types.  
For example, although 80\%$\pm 6$\% of the BLAGNs in our sample have 
\Geff~$\ge 1.2$, so do 36\%$\pm 6$\% of our HEX sources and 30\%$\pm
5$\% of our absorbers and star formers.  (Of course, this also means
that 20\%$\pm 3$\% of the BLAGNs in our sample have \Geff~$<1.2$.)
Even considering a more extreme X-ray softness cut-off, we still 
find that 12\%$\pm 3$\% of the HEX sources and 6\%$\pm 2$\% of the
absorbers and star formers in our sample have \Geff~$\ge 1.7$.
A number of authors have suggested possible ways to account
for this overlap through observational problems.  In this section 
we briefly consider these and conclude that they are 
not major contributors to the overlap.

X-ray spectral variability has been suggested as a possible 
explanation for mismatches between X-ray and optical 
classifications of AGNs \citep{paolillo04}. Since the
X-ray and optical observations of the sources in our 
sample were not obtained simultaneously, we need to consider
this possibility.  \citet{yang04} tested a subsample of 
CLASXS sources for flux variability. Of the 60 sources with 
$4\times10^{-14}<f_{0.4-8~\rm keV}<8\times10^{-14}$~erg~cm$^{-2}$~s$^{-1}$, 
70\% showed flux variability over a period of days to a year, 
depending on the location of the source. The change in spectral 
slope for 20\% of these sources was sufficient to transform 
the X-ray spectral type from X-ray soft (\Geff~$\ge1.2$) 
to X-ray hard (\Geff~$<1.2$) or vice-versa. However, of the 60 sources 
tested for variability, only two are BLAGNs (CLASXS~\#293 and 
CLASXS~\#397), and neither of these exhibits sufficient spectral 
variability to change its X-ray spectral type. 
Of the 58 other sources tested for variability, two fall within our 
sample of non-BLAGNs with \Geff~$\ge 1.2$ (CLASXS~\#286 and 
CLASXS~\#199). CLASXS~\#286 does not exhibit sufficient spectral 
variability to change its X-ray spectral type, while CLASXS~\#199 
does.  We therefore conclude that while X-ray spectral variability 
does seem to be able to account for some of the X-ray unabsorbed
(\Geff~$\ge 1.2$), optically obscured (non-BLAGN) sources, it is 
unlikely that it can account for the majority.

\citet{silverman05} argued that the broad emission lines for 
their X-ray unabsorbed, optically obscured sources 
were redshifted out of their optical spectral window. 
However, of the 84 X-ray unabsorbed, optically obscured sources in our
spectroscopically observed
$2-8~\rm keV$ sample, all but one have at least one of the 
potentially broadened emission lines (H$\alpha$, H$\beta$, MgII, 
Ly$\alpha$, or CIV) measurable within the spectral window. Thus, 
this explanation also cannot account for the majority of our X-ray
unabsorbed, optically obscured sources.

\citet{moran02} suggested that the light from the host galaxy 
may dilute the AGN signal. They found that 60\% of their sample 
of local Seyfert~2s (i.e., HEX sources) would be classified as 
optically normal (i.e., as absorbers or star formers) if only the 
total emission were available (as would be the case for observations 
of high-redshift galaxies; see also \citealt{cardamone07}). 
Following this suggestion,
\citet{severgnini03}, \citet{hasinger05}, and \citet{garcet07} 
posited that the absorbers and star formers in their sample were
BLAGNs (rather than HEX sources) whose light was being diluted by 
the host galaxy light. 
However, the only possible evidence for this comes from one of
the two sources in \citet{severgnini03}. \citet{severgnini03}
obtained improved optical 
spectroscopic observations of two X-ray unabsorbed, optically
normal galaxies. Previous optical spectra for
these sources did not cover the H$\alpha$ line. As a result
of the higher signal-to-noise and optimal wavelength coverage of 
new optical spectra they obtained, they found that the two X-ray unabsorbed 
sources exhibited strong and (possibly) broad H$\alpha$. 
They then simulated the effect of placing each source at a higher 
redshift using a template spectrum with both an AGN and a host galaxy 
component and concluded that for one of the sources, if it had instead 
been at $z>0.2$, the H$\alpha$ line would not have been 
visible due to dilution by the host galaxy light. 

We stress that no evidence was presented by either
\citet{hasinger05} or \citet{garcet07} to show that their optically
normal sources 
were diluted BLAGNs rather than diluted HEX sources.
Furthermore, as discussed in the Introduction, \citet{cowie09} 
found that the non-BLAGNs in our OPTX sample with $0.9<z<1.4$ are 
UV faint (as opposed to the BLAGNs, which are UV bright), indicating 
that we are not misclassifying as non-BLAGNs sources that are 
really BLAGNs. Similarly, \citet{barger05} argued against
misclassification being a problem based on the weakness of the 
UV nuclei in the non-BLAGNs relative to their X-ray light.
Thus, host galaxy dilution does not appear to be the
explanation for the X-ray unabsorbed, optically 
normal sources in our sample.

Moreover, it does not seem likely that host galaxy dilution could 
cause BLAGNs to appear as HEX sources (i.e., wash 
out the broad lines while still allowing the narrow lines to be
observed). The composite UV-optical spectrum for the sample of 
AGNs and quasars in the Large Bright Quasar Survey revealed that 
the equivalent widths of the broad lines is significantly greater 
than that of the narrow lines \citep{francis91}. In fact, in 
luminous QSOs there are often no narrow lines at all \citep{zheng97}. 
Thus, host galaxy dilution also does not appear to be the
explanation for the X-ray unabsorbed, HEX sources in our sample.

\section{Conclusions}
\label{conclusions}

We find that the observational problems discussed in
Section~\ref{discussion}
cannot explain most of the overlap that we see in the X-ray 
spectral properties for different optical spectral types.
Until a better understanding is reached for how the X-ray 
and optical classifications relate to the obscuration of the 
central engine, the use of a mixed classification 
scheme can only complicate the interpretation of X-ray AGN samples.  
As a case in point, \citet{cowie09} found that a number of the BLAGNs 
in our OPTX sample have high ratios of ionizing flux to X-ray flux
yet low values of \Geff. This suggests that there is not 
a one-to-one correspondence between the X-ray photon index and 
the opacity of the source, since any substantial neutral hydrogen 
opacity ($N_{H}>3\times10^{17}$~cm$^{-2}$) would absorb the 
ionizing flux, as \citet{cowie09} observed for the OPTX non-BLAGNs. 
Finally, we emphasize that any classification scheme which uses 
X-ray hardness ratio or, equivalently, effective photon index 
will be highly redshift dependent, which can introduce serious redshift
bias.  On the basis of this study, we advocate the adoption of 
a pure optical classification scheme for studying AGN with low
signal-to-noise X-ray spectra. However if high quality X-ray spectra
are available, they form an equally valid method of categorizing the
objects \citep[e.g.][]{winter09}.

\acknowledgements

L.~T.~was supported by a National Science Foundation Graduate Research
Fellowship and a Wisconsin Space Grant Consortium Graduate Fellowship
Award during portions of this work. We also gratefully acknowledge 
support from NSF grants AST 0407374 and AST 0709356 (L.~L.~C.) and 
AST 0239425 and AST 0708793 (A.~J.~B.), the University of
Wisconsin Research Committee with funds granted by the Wisconsin
Alumni Research Foundation and the David and Lucile
Packard Foundation (A.~J.~B.). A.~J.~B.~thanks the Aspen Center for
Physics for hospitality during the completion of this work. This
article is part of L.~T.'s Ph.D.~thesis work at the University of
Wisconsin-Madison. 


\end{document}